\documentclass[a4paper,11pt]{article}

\usepackage[colorlinks=false,citecolor=blue,urlcolor=black]{hyperref}
\usepackage{bm}
\usepackage{bbm}
\usepackage{amssymb}
\usepackage{amsmath}
\usepackage{amsthm}
\usepackage{mathrsfs}
\usepackage{mathtools}
\usepackage{dsfont}
\usepackage{thmtools} 
\usepackage{thm-restate}
\usepackage{booktabs}
\usepackage{multirow}
\usepackage{diagbox}
\usepackage{adjustbox}
\usepackage{xcolor}
\usepackage{authblk}
\usepackage[natbib,doi=false,isbn=false,url=false,eprint=false, 
            style=nature]{biblatex}
\usepackage{url}
\usepackage{appendix}
\usepackage[onehalfspacing]{setspace}
\usepackage[margin=1in]{geometry}
\usepackage{empheq}  
\usepackage{float}

\usepackage{enumerate}

\AtEveryBibitem{\clearfield{month}} 

\newcommand{\Dconvergence}{\xlongrightarrow{d}}

\begin{document}

\def\spacingset#1{\renewcommand{\baselinestretch}%
{#1}\small\normalsize} 


\renewcommand*{\Affilfont}{\normalfont\itshape\small} 

\title{\centering{Numerical models outperform AI weather forecasts of record-breaking extremes}}

\author[a,b,*]{Zhongwei~Zhang}
\author[c]{Erich~Fischer}
\author[d,e]{Jakob~Zscheischler}
\author[a,*]{Sebastian~Engelke}

\affil[a]{Research Institute for Statistics and Information Science, University of Geneva, Geneva, Switzerland}
\affil[b]{Scientific Computing Center, Karlsruhe Institute of Technology, Karlsruhe, Germany}
\affil[c]{Institute for Atmospheric and Climate Science, Department of Environmental Systems Science, ETH Zurich, Zurich, Switzerland}
\affil[d]{Department of Compound Environmental Risks,\par Helmholtz~Centre~for~Environmental~Research~--~UFZ, Leipzig, Germany}
\affil[e]{Department of Hydro Sciences, TUD Dresden University of Technology, Dresden, Germany}
\affil[*]{Corresponding authors: zhongwei.zhangss@gmail.com, sebastian.engelke@unige.ch}

\date{}
\maketitle

\medskip

\begin{abstract}
Artificial intelligence (AI)-based models are revolutionizing weather forecasting and have surpassed leading numerical weather prediction systems on various benchmark tasks. However, their ability to extrapolate and reliably forecast unprecedented extreme events remains unclear.
Here, we show that for record-breaking weather extremes, the numerical model High RESolution forecast (HRES) from the European Centre for Medium-Range Weather Forecasts still consistently outperforms state-of-the-art AI models GraphCast, GraphCast operational, Pangu-Weather, Pangu-Weather operational, and Fuxi. 
We demonstrate that forecast errors in AI models are consistently larger for record-breaking heat, cold, and wind than in HRES across nearly all lead times. 
We further find that the examined AI models tend to underestimate both the frequency and intensity of record-breaking events, and they underpredict hot records and overestimate cold records with growing errors for larger record exceedance.
Our findings underscore the current limitations of AI weather models in extrapolating beyond their training domain and in forecasting the potentially most impactful record-breaking weather events that are particularly frequent in a rapidly warming climate. Further rigorous verification and model development is needed before these models can be solely relied upon for high-stakes applications such as early warning systems and disaster management.
\end{abstract}


\begin{refsection}

\section*{Introduction}
Record-breaking weather extremes, such as the 2021 Pacific Northwest, 2010 Russian and 2003 European heatwaves, and winter storms Lothar in 1999 and Kyrill in 2007, have caused numerous fatalities and severe impacts on society, the economy, and ecosystems~\citep{White2023,Barriopedro2011,GarciaHerrera2010,Wernli2002,Fink2009}. The level of disaster preparedness and adaptation to extreme events is strongly influenced by events observed in recent decades. Consequently, after extended periods without major events, or when events substantially exceed previous record levels, socio-economic impacts tend to be particularly large. 
 
In addition to long-term disaster preparedness~\citep{Kelder2025}, accurate numerical weather prediction (NWP) is critical for early-warning systems to save lives and reduce the impacts of climate extremes~\citep{Bauer2015}. Recently, a new generation of AI weather models has reached and sometimes exceeded forecast skills of state-of-the-art NWP systems \citep{Bi2023,Lam2023,Lang2024,Bodnar2025}. These models offer considerable advantages in speed and energy efficiency, raising important questions about their potential to supplement or eventually replace traditional NWP systems~\citep{Schultz2021}. 

Before warnings for population and critical infrastructure are routinely based on AI models, their performance needs to be further evaluated. 
In particular, their reliability in forecasting extreme events remains less well understood.
Such events are, by definition, rare and contribute little to aggregated overall skill metrics \citep{Watson2022}. Nevertheless, recent studies suggest that AI models perform well---and in some cases even better than numerical models---in forecasting extreme weather events \citep{BenBouallegue2024, Olivetti2024}, particularly for longer lead times \citep{Lam2023,Bi2023}.

Current forecast evaluation approaches for extreme events typically focus on extreme events exceeding a certain threshold for one or several given variables, such as extreme wind speeds \citep{Olivetti2024}, tropical cyclones \citep{Bi2023, Lam2023, BenBouallegue2024}, and high and low temperatures \citep{Lam2023, BenBouallegue2024, Olivetti2024, Meng2025}. However, due to small sample sizes, the thresholds are often set to, say, the 95th percentile of the test data, thus capturing mostly moderate extremes. 
Much less is known about record-breaking events, a subset of extreme events that are unprecedented in the observational record.
Given the current high rate of global warming, record-breaking events sometimes exceed previous record levels by large margins and have been referred to as black or gray swans~\citep{Sun2025}, or record-shattering extremes~\citep{Fischer2021,Fischer2023}. 

A number of case studies have shown mixed results on the ability of AI weather models to extrapolate beyond the range of their training data.
For instance, a seasonal AI forecasting model \citep{WattMeyer2024}
struggled to predict North Atlantic Oscillation values that extended outside its training distribution \citep{Kent2025}. 
While AI models appear to outperform traditional NWP models on tracking tropical cyclones \citep{pathak2022fourcastnet, Bi2023, Lam2023}, they tend to underpredict the intensity of the most extreme storms, as measured by mean sea-level pressure \citep{demaria2024, charlton2024ai, Sun2025}.
Similar limitations in reaching unprecedented amplitudes have also been observed in other high-impact events such as heatwaves, winter storms, or compound extremes \citep{Pasche2025}.
On the other hand, the unprecedented 2024 rainfall in Dubai was well predicted by GraphCast, suggesting that generalization to new events may be possible if they share dynamical similarity with past extremes from other regions \citep{Sun2025a}.

\begin{figure}[ht!]
    \centering
    \includegraphics[width=0.95\textwidth]{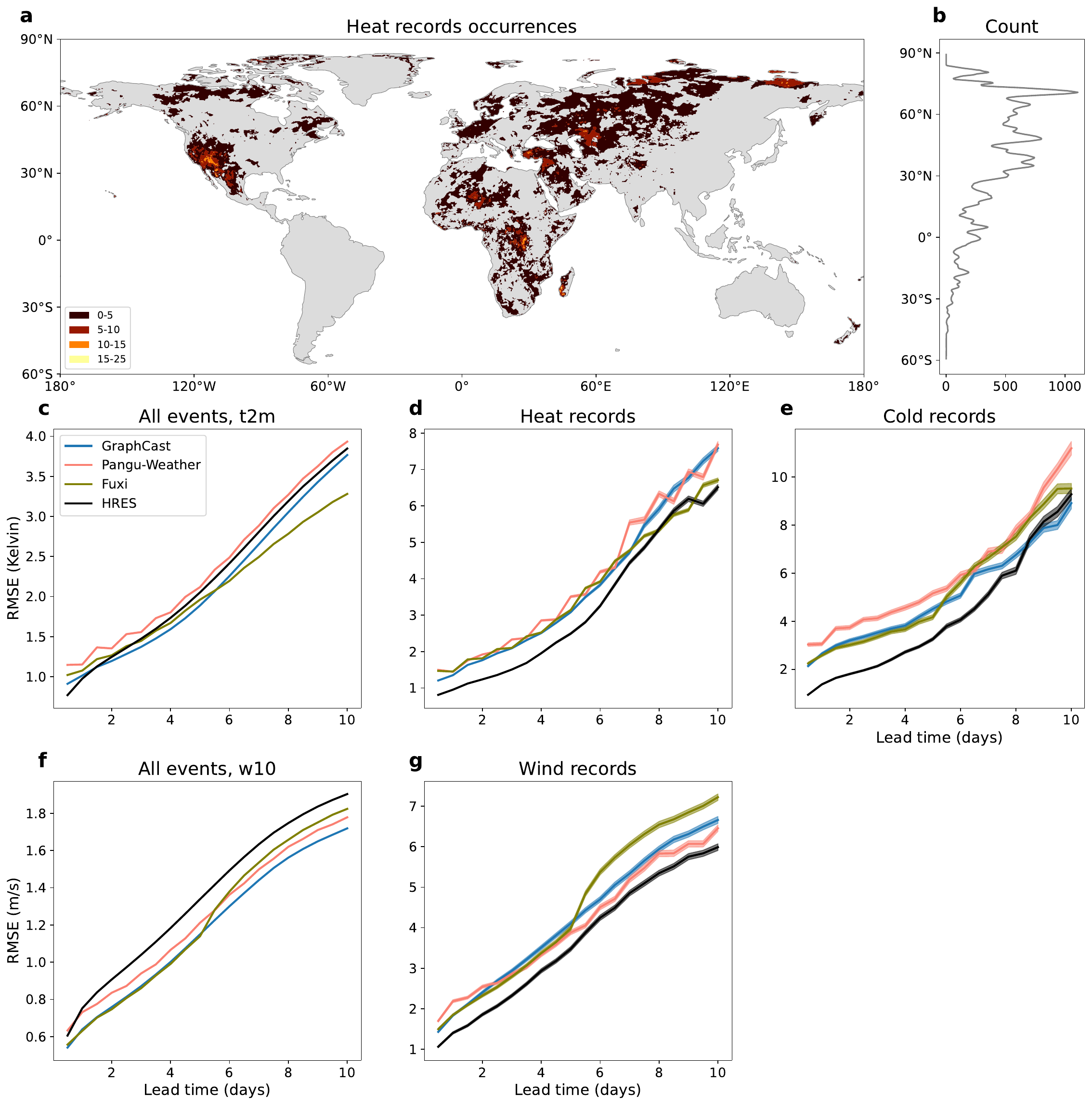}
    \caption{\textbf{Model performance on all events and record-breaking events.} \textbf{a}, Number of heat records in 2020 in ERA5. \textbf{b}, Number of heat records per latitude. \textbf{c}--\textbf{g}, Root mean square error (RMSE) of forecasted 2m temperature and 10m wind speed over land (excluding the Antarctic region) of HRES, Pangu-Weather, GraphCast, and Fuxi for all events (\textbf{c, f}) and only record-breaking events (\textbf{d, e, g}) in 2020 for different lead times. The transparent shaded areas indicate $95\%$ confidence bands. 
    }
    \label{figure::RMSE_t2m_w10_2020}
\end{figure}

However, these insights primarily rely on isolated case studies of specific events, whose conclusions are inherently difficult to generalize due to the unique features of the analyzed events and models.
To systematically evaluate extrapolation in state-of-the-art AI weather models, we construct a benchmark dataset consisting of record-breaking events for heat, cold, and wind extremes. This dataset includes all observations during the test years 2018 and 2020 that exceed the respective historical records from the training period 1979--2017 of all considered AI models. For each variable, a record-breaking event is defined locally, per grid cell and per calendar month, yielding a large sample size even in individual years (see Methods). For the year 2020, this yields $162,751$ heat, $32,991$ cold, and $53,345$ wind records, which are spread across different seasons and climatic zones from tropics to high latitudes (Fig.~\ref{figure::RMSE_t2m_w10_2020}a,b, and Supplementary Fig.~\ref{figure::map_records_ERA5_2020}a--d). 
The dataset includes many prominent record-breaking events, such as the Siberian heatwave in early 2020~\citep{Overland2020} and the U.S.~heatwave of August 2020~\citep{Li2023}.
Evaluating AI models on this record dataset challenges them to forecast on out-of-distribution data, which is known to be difficult for neural networks in the machine learning literature.

We assess the extrapolation performance on our benchmark dataset of record-breaking events of three leading deterministic AI weather models: GraphCast \citep{Lam2023}, Pangu-Weather \citep{Bi2023}, Fuxi \citep{Chen2023}, as well as the operational variants of GraphCast and Pangu-Weather. Their performance is compared to HRES from the European Centre for Medium-Range Weather Forecasts (ECMWF), which is widely considered as the leading NWP model.

\section*{Model comparison on records’ intensity}
Consistent with previous studies \citep{Lam2023, Bi2023, Chen2023, Rasp2024}, we find that, on overall performance, all AI models---except Pangu-Weather---outperform the ECMWF model HRES in forecasting 2-meter temperature across most lead times (Fig.~\ref{figure::RMSE_t2m_w10_2020}c). Forecast accuracy is quantified using root mean square errors (RMSE), computed over all 00 and 12 UTC time steps in test year 2020 and over all land grid points (excluding the Antarctic region; see Methods). For 10-meter wind speed, all AI models consistently outperform HRES across nearly all lead times (Fig.~\ref{figure::RMSE_t2m_w10_2020}f).

However, the predictive skill is drastically different for record-breaking temperature and wind events in 2020. Restricting the RMSE to record-breaking events, the numerical HRES model consistently outperforms all AI models for hot and cold temperature records as well as wind speed records across almost all lead times (Fig.~\ref{figure::RMSE_t2m_w10_2020}d,e,g). The performance gap is most pronounced for short lead times. For lead times beyond 5 days HRES still generally performs better but to a lesser extent. This aligns with previous findings that AI models tend to perform relatively better at longer lead times~\citep{Lam2023}. 

While, due to limited data availability, the evaluation is shown for a single year only as in most previous studies~\citep{Bi2023,Olivetti2024}, we observe the same pattern in 2018, the other year for which forecasts are available, except for Fuxi (Supplementary Fig.~\ref{figure::RMSE_t2m_w10_2018}). The years 2018 and 2020 are distinctly different in terms of ENSO conditions,
with 2018 transitioning from La Nina to El Nino and 2020 undergoing a strong El Nino to La Nina shift.
Since ENSO strongly influences the occurrence of temperature records~\citep{Fischer2025}, particularly in the tropics, the consistent outperformance of HRES across both years shows the robustness of the results. The better skill of HRES in predicting record-breaking events is further consistent across different seasons and a wide range of different climate zones, including tropics, subtropics, mid-latitudes and northern high latitudes (Fig.~\ref{figure::RMSE_t2m_w10_2020}a, Supplementary Figs.~\ref{figure::RMSE_regional_2020} and \ref{figure::RMSE_seasonal_2020}).

While it is common to evaluate ERA5-trained AI models against ERA5 reanalysis, and HRES against its own analysis at lead time 0 (HRES-fc0)~\citep{Bi2023,Lam2023,Chen2023} (see Methods), this approach can complicate comparisons due to different horizontal resolution: ERA5 has a resolution of $0.25^\circ$, whereas HRES operates at $0.1^\circ$.
To assess the sensitivity of our findings to the choice of different reference datasets, we also evaluate operational versions of GraphCast and Pangu-Weather against HRES on a common test dataset of record-breaking events identified using HRES-fc0 as observational ground truth. Also in this setting, HRES consistently outperforms the AI models on the records (Supplementary Fig.~\ref{figure::RMSE_t2m_w10_2020_oper}). 

Selecting a subset of extreme events based on observations can favor models that produce too many extreme forecasts---a problem known as the forecaster's dilemma \citep{Lerch2017} (see Methods for discussion). Thus, we construct an alternative benchmark avoiding the forecaster's dilemma, based on events where the forecast itself, rather than the observation, exceeds the training record \citep{Holzmann2014}. Results from this forecast-conditioned evaluation (Supplementary Fig.~\ref{figure::RMSE_forecast_bias_2020_oper_cond_HRES_GC}) are consistent with the previous conclusion that HRES outperforms current AI models in forecasting records.

\begin{figure}[ht!]
    \centering
    \includegraphics[width=0.95\textwidth]{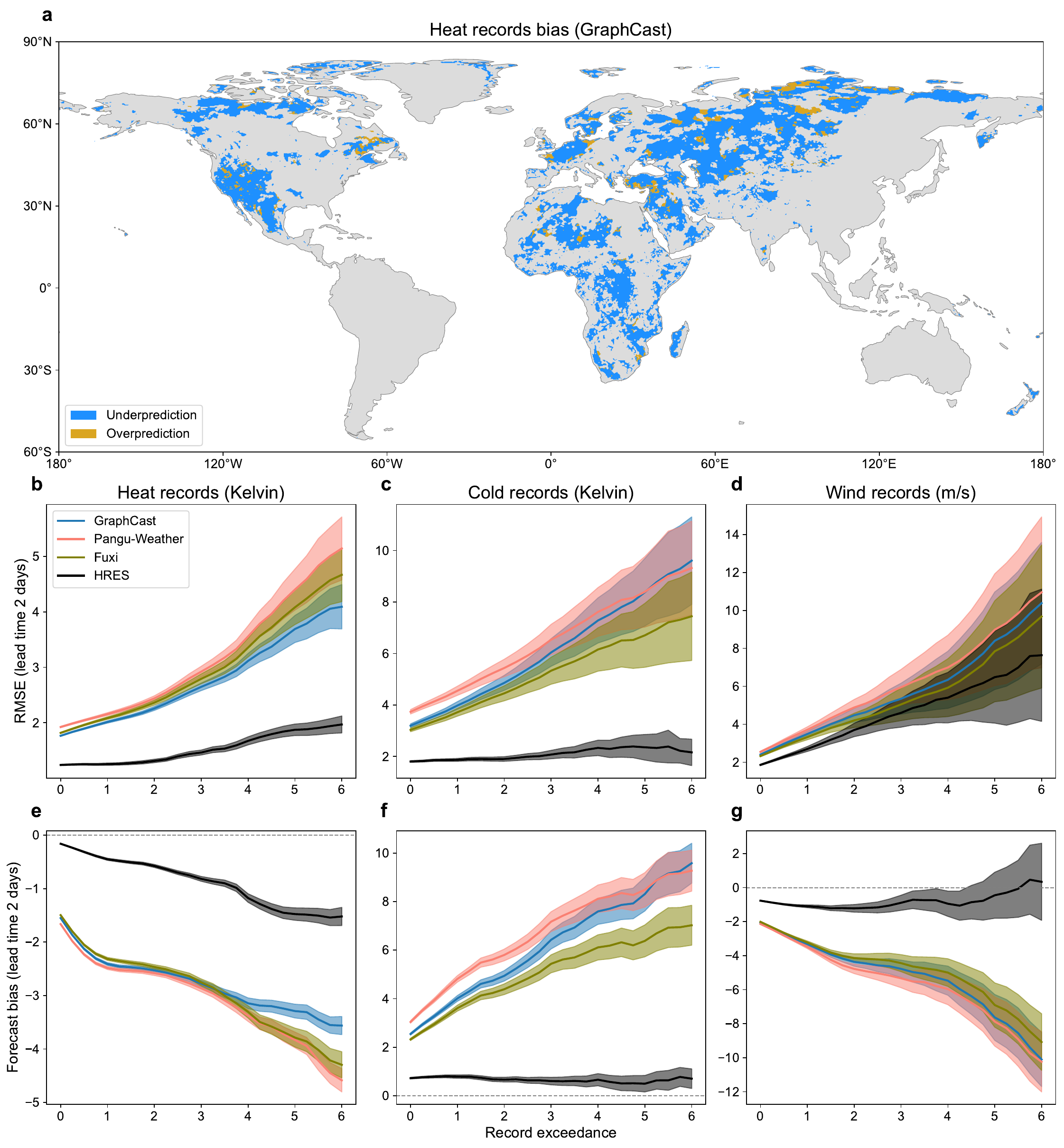}
    \caption{\textbf{Forecast bias against record exceedance.} \textbf{a}, Forecast bias of the maximum heat records (GraphCast).  \textbf{b}--\textbf{d}, RMSE of 2m temperature for heat and cold records, and 10m wind speed for wind records for events in 2020 that exceed the record by at least a certain margin (x-axis). Only land pixels (excluding the Antarctic region) are considered. \textbf{e}--\textbf{g}, Forecast bias of heat, cold, and wind records, for events that exceed the record by at least a certain margin. The transparent shaded areas indicate $95\%$ confidence bands. 
    }
    \label{figure::forecast_bias_record_exceedance}
\end{figure}

\section*{AI models underestimate intensities of records}

While we demonstrate that AI models underperform compared to HRES in forecasting record-breaking events, their errors may arise from  over- or underprediction of event intensity. 
When considering all data of the test year 2020, all models have relatively small, unsystematic biases (Supplementary Fig.~\ref{figure::FB_all_records_2020}a,d).
To better understand model behavior beyond their training domain, we compare forecast accuracy and bias against the record exceedance, that is, the margin by which a record is exceeded. 
We find that AI models generally underpredict temperature during high records and overpredict during low records.
This pattern is shown for GraphCast and heat records (Fig.~\ref{figure::forecast_bias_record_exceedance}a). The systematic underprediction is remarkably consistent across regions, seasons, and location in tropics, subtropics and mid- to high-latitudes, despite the fact that the physical drivers of heat records vary substantially across regions. 
This behavior is not limited to a single model: other AI models show similar patterns of intensity underestimation, while HRES demonstrates a more balanced distribution of over- and underpredictions (Supplementary Fig.~\ref{figure::map_under_over_prediction_2020_non_oper}). These results strongly suggest that AI model forecast errors are at least partly due to systematic extrapolation limitations.

For all record types, the errors of the three AI models seem to grow almost linearly with respect to the degree of record exceedance (Fig.~\ref{figure::forecast_bias_record_exceedance}b--d for a lead time of 2 days; additional lead times in Supplementary Fig.~\ref{figure::RMSE_record_exceedance_2020}).
This trend indicates that forecast bias is the primary driver of error
(Fig.~\ref{figure::forecast_bias_record_exceedance}e--g and Supplementary Fig.~\ref{figure::FB_all_records_2020}): the greater the record exceedance, the larger the underestimation of event intensity. The models behave as if their predictions have an implicit (soft) cap at a certain local value. 
In contrast, the physical HRES model is more robust to extreme records exceedances. For temperature records, HRES exhibits a nearly constant error across increasing exceedances. For wind records, it shows a mild tendency of underestimation, though far less so than AI models. Overall, HRES exhibits lower forecast bias for all records types, and bias is not the dominant source of error, particularly for cold and wind records. 

Importantly, this behavior, shown here for the evaluation year 2020, is fully consistent with results from both the operational forecasts in 2020 and non-operational forecasts in 2018 (Supplementary Figs.~\ref{figure::RMSE_record_exceedance_operational} and~\ref{figure::RMSE_record_exceedance_2018}).
The systematic, one-sided bias observed across event types, lead times, regions and independent years provides strong evidence that current AI models have a structural extrapolation problem when forecasting record-breaking events.

\begin{figure}[ht!]
    \centering
    \includegraphics[width=\textwidth]{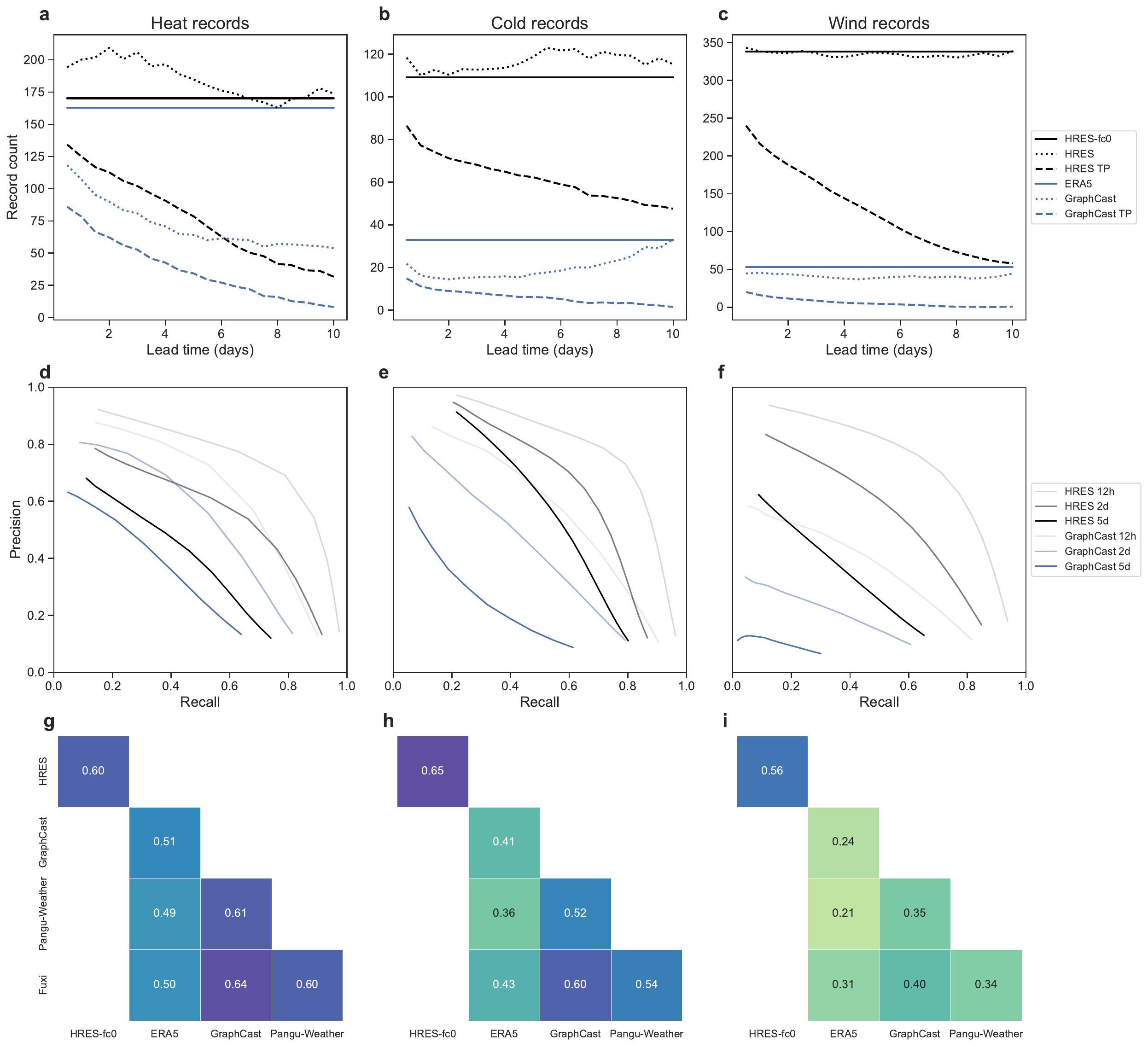}
    \caption{\textbf{Prediction of occurrence of record-breaking events.} \textbf{a}--\textbf{c}, Counts (in thousands) of heat, cold, and wind records in the ground truth ERA5 and HRES-fc0 data, and GraphCast and HRES forecast data for 2020, as well as counts of their true positives (TP) over land (excluding the Antarctic region). \textbf{d}--\textbf{f}, Precision and recall curves of GraphCast and HRES forecasts when the records are used as the threshold for different lead times. 
    \textbf{g}--\textbf{i}, Correlations between the indicator functions of whether the ground truth or 2-day forecasts exceed the record. 
    }
    \label{figure::precision_recall_GraphCast}
\end{figure}

\section*{Model comparison on records’ occurrence}

We further test the ability of AI models to predict not only the intensity but also the frequency of record-breaking events. We find that, in addition to underestimating event intensity, AI models systematically underpredict the number of records relative to their ERA5 ground truth (Fig.~\ref{figure::precision_recall_GraphCast}a--c). This underestimation results in a high number of false negatives and consequently low recall (defined as the ratio of true positives to the observed positives). In contrast, HRES forecasts a number of records comparable to its HRES-fc0 ground truth, with a slight overestimation for heat records at smaller lead times.

Correctly predicting the number of record-breaking events does not imply accurate timing. In risk management, the trade-off between false positives and false negatives is typically evaluated using precision-recall curves (see Methods). 
Across all record types and lead times, HRES's precision-recall curves are consistently better than GraphCast's, in the sense that they are closer to the ideal point (precision $=$ 1, recall $=$ 1), indicating superior classification performance for heat, cold and wind records (Fig.~\ref{figure::precision_recall_GraphCast}d--f). This is in contrast with earlier results that demonstrate that GraphCast outperforms numerical models for more moderate extreme events~\citep{Lam2023}.
Similar results are observed for Pangu-Weather and Fuxi, where HRES again shows a better classification skill across all lead times (Supplementary Figs.~\ref{figure::precision_recall_Pangu_2020_non_oper} and~\ref{figure::precision_recall_Fuxi_2020_non_oper}).

As an additional evaluation, we convert both forecast and ground truth into binary variables (1 if a record is exceeded and 0 otherwise) and compute the correlation between them (see Methods).
This metric complements the precision-recall analysis by incorporating true negatives and measuring the degree of dependence between different models' forecasts.
HRES has a higher correlation with its ground truth HRES-fc0 than the AI models with their ground truth ERA5, reaffirming its superior performance in forecasting record-breaking events (Fig.~\ref{figure::precision_recall_GraphCast}g--i). Interestingly, all AI models are positively correlated with each other, showing that they tend to make errors on the same events. This may be due to shared biases learned from their common training data.

\section*{Discussion}

Our findings consistently show that current AI models underperform HRES in forecasting record-breaking events. 
They tend to underpredict heat and wind speed records, and overpredict cold records, with greater forecast biases the larger the record margin. This strongly suggests a systematic extrapolation problem in these models.

All current state-of-the-art AI weather models are built on neural network architectures such as transformers \citep{Chen2023, Bi2023} or graph neural networks \citep{Lam2023, Price2024, Lang2024}. In machine learning, extrapolation, also referred to as out-of-distribution generalization, is a well-known fundamental challenge in these models. It has been observed in a range of applications, including image classification \cite{Ganin2016}, protein fitness prediction \citep{Freschlin2024}, and large language models \citep{Hupkes2023}. Our record benchmark dataset
is explicitly designed to test this out-of-distribution problem within AI weather models (see Methods for discussion).

The AI models studied here do not use any knowledge of physical principles and do not explicitly enforce energy balances or other physical constraints \citep{sel2023, bon2024}. They are purely data-driven and essentially interpolate between observed historical weather patterns in the training period 1979--2017 to produce forecasts for new initial conditions in the test period. This is in stark contrast to physics-based numerical models like HRES that strongly rely on partial differential equations describing the evolution of the atmosphere based on our understanding of physics.
This fundamental difference in modeling philosophy likely explains the discrepancy in performance between AI and NWP models for record-breaking events  (Fig.~\ref{figure::RMSE_t2m_w10_2020}c--g).
While AI models excel when the test set closely resembles the training distribution, capturing complex atmospheric patterns and improving skill on average conditions, they struggle when forecasting unprecedented events outside the training domain, even at short lead times.
The nearly linear increase of the biases with record exceedance (Fig.~\ref{figure::forecast_bias_record_exceedance}e--g) suggests an implicit cap in AI forecasts around the most extreme training observation. Physical models do not have such a bound since physical principles allow them to extrapolate, and, consequently, they exhibit less bias across record magnitudes. Moreover, deterministic AI weather forecasts often smooth out fine-scale spatial features such as sharp wind peaks. By contrast, recent probabilistic AI weather models \citep{Price2024, Lang2024_prob} aim to preserve variability and avoid such smoothing. Still, as our results suggest, even these models likely face similar extrapolation challenges when forecasting out-of-distribution, record-breaking events.

Several promising avenues exist to address this shortcoming in future generations of AI weather models.
One strategy is data augmentation, a widely used technique in machine learning to improve robustness to unseen scenarios by enriching the training data \citep{shorten2019survey}. In weather and climate modeling, a key advantage is that numerical climate models can produce very large amounts of physically plausible extreme events outside the training domain. Augmenting training with simulations from different climate regimes \citep{Bodnar2025} or record-breaking events from ensemble boosting \citep{Fischer2023} could allow AI models to learn from more extreme events than in the original training data. This approach has already shown promise: FourCastNet's \citep{pathak2022fourcastnet} performance on tropical cyclones improves significantly when trained on datasets that include such events \citep{Sun2025}. 
Another promising direction involves hybrid modeling, where specific parameterizations in physical climate models are replaced with AI components~\citep{Shaw2025}. These models combine the efficiency and learning capacity of AI models with the physical consistency and extrapolation ability of physical models. 
Hybrid models such as NeuralGCM \cite{kochkov2024neural} remain fully differentiable and thus allow for efficient optimization of initial conditions \cite{Whittaker2025}, for instance.
Finally, the loss functions used to train AI weather models are typically designed to predict the mean or bulk of the distribution. To improve extrapolation performance on extremes, it may be possible to adapt principles from statistical learning and extreme value theory \citep{Shen2024, buritica024,bou2022}.

Given the remarkably fast evolution of AI models in recent years, there are promising ways to further improve these models even for forecasting record-breaking extremes that will continue to frequently occur in a rapidly warming climate.
Nevertheless, the current generation still underperforms HRES exactly during the potentially most impactful weather events, including record-breaking heat and cold events as well as wind storms. Thus, it remains vital to fund and run NWP and AI weather models in parallel and to rigorously evaluate their performance for the most impactful type of weather events.



\printbibliography
\end{refsection}

\begin{refsection}

\newpage
\section*{Methods}

\subsection*{Models and data}

For the definition of records we use the ECMWF's ERA5 reanalysis data~\citep{Hersbach2020} from 1979--2017 with daily observations at 00, 06, 12, and 18 UTC time. This dataset coincides with the training data of almost all AI models considered in this paper. 
The time points in this training data are denoted by $T_{\text{train}}$.
The ERA5 data is available on a $0.25^\circ \times 0.25^\circ$ latitude-longitude grid. Throughout the paper we only consider data over land. We use the land-sea mask from the ERA5
and follow ECMWF~\citep{Owens2018} by defining a grid cell as land if more than $50\%$ of the cell is covered by land; otherwise it is considered as sea.
We exclude the Antarctic region (grid cells with latitude in the range $(-60^\circ, -90^\circ]$) due to aberrant behavior exhibited by some AI models in this region, and denote the remaining set of land grid cells ($244,450$ grid cells in total) from the ERA5 dataset by $G_{0.25^\circ}$.

We use forecasts from the state-of-the-art AI models GraphCast~\citep{Lam2023}, Pangu-Weather~\citep{Bi2023}, and Fuxi~\citep{Chen2023} from a test period $T_{\text{test}}$, which is either of the years 2018 or 2020 in our analyses.
For the same period, we use forecasts from the High RESolution model (HRES) of ECMWF for comparison.
All the forecast data are publicly available from WeatherBench~2~\citep{Rasp2024}.
Pangu-Weather and Fuxi are trained and validated on ERA5 data from 1979--2017; the GraphCast forecast data for years 2018 and 2020 are produced by two slightly different versions of GraphCast, i.e., the 2018 data are generated by the GraphCast model trained on ERA5 data from 1979--2017, whilst the 2020 data are generated by the GraphCast model trained with ERA5 data from a slightly extended period 1979--2019.
In addition, we also employ the operational versions of GraphCast and Pangu-Weather. The former has been fine-tuned on the HRES-fc0 data from 2016--2021, while the latter was used in an operational setting without fine-tuning.

As ground truth for the AI models we use ERA5 data with locations in $G_{0.25^\circ}$ in the test period. For HRES and the operational AI models we use HRES-fc0 as ground truth. 
Using these two different datasets to evaluate the forecasts against is the standard approach in the literature of AI weather models to avoid unfair comparisons~\citep{Bi2023,Lam2023,Chen2023}.

\subsection*{A benchmark dataset of record-breaking events}

To define a dataset of record-breaking events in a given year (e.g., 2020) for a variable $x$ of interest (e.g., 2-meter temperature), we first compute the corresponding record in the ERA5 data $T_{\text{train}}$ in the training period of the AI models from 1979--2017.
We specify whether we consider records in the positive direction (e.g., heat records) or the negative direction (e.g., cold records) by superscripts max or min, respectively.
A record $r_{s,m}^{x,\max}$ of variable $x$ is defined locally per grid cell $s\in G_{0.25^\circ}$ and per month $m\in\{\text{January}, \dots, \text{December}\}$.
More precisely, we define 
\begin{align}\label{record_def}
    r_{s,m}^{x,\max} = \max_{t \in T_{\text{train}}; t\in m} x_{s,t}, 
\end{align}
where $x_{s,t}$ is the value of variable $x$ at location $s$ and time $t$, and $t\in m$ indicates that only time points in month $m$ are considered. 

We define the set $R^{x,\max}\subseteq G_{0.25^\circ}\times T_{\text{test}}$ of record-breaking events of variable $x$ consisting of location-time pairs encoding where and when the event occurred. The test period $T_{\text{test}}$ contains all time points at 00 and 12 UTC in the test year, i.e., the year 2018 or 2020 in our analyses.
We denote by $m(t)$ the month corresponding to a time $t\in T_{\text{test}}$, so that 
$x_{s,t} > r_{s,m(t)}^{x,\max}$ means that observation $x_{s,t}$ exceeds its respective
monthly historical record. With this we have
\begin{align}\label{R_def}
    R^{x,\max} &= \{(s,t)\in G_{0.25^\circ}\times T_{\text{test}}: x_{s,t} > r_{s,m(t)}^{x,\max}\}. 
\end{align}

We do not evaluate forecasts initiated at 06 and 18 UTC since the HRES forecasts with these initializations are only available for 3.75 days at ECMWF, and all AI-based forecasts are only available for initializations at 00 and 12 UTC on WeatherBench~2~\citep{Rasp2024}.
In addition, we only consider lead times that are multiples of 12 hours to ensure that the subsets of ERA5 data used as input and ground truth for AI models have the same +9h lookahead~\citep{Lam2023} (ERA5 and HRES-fc0 have different data assimilation windows: ERA5 has +3h lookahead at 06 and 18 UTC and +9h loohahead at 00 and 12 UTC, while HRES-fc0 have +3h lookahead for all four time points).
Consequently, this comparison setup disadvantages HRES due to the mismatch between a +9h lookahead of ERA5 input and +3h lookahead of HRES-fc0 input, thereby strengthening our main result that HRES outperforms AI models on record-breaking events.

Note that the notion of a record-breaking event is to be understood relative to the training period. We do not update the record if a larger event has occurred after 2017 (the end of the training period).
The reason is that AI models are not retrained and a record-breaking event in the test period will not inform or improve the model for later time steps.

Using as test data the ERA5 ground truth in 2020 we obtain 162,751 records for heat, 32,991 for cold and 53,345 for wind; see the geographical distribution of these records in the map in Fig.~\ref{figure::RMSE_t2m_w10_2020}a and Supplementary Fig.~\ref{figure::map_records_ERA5_2020}, respectively.
For the analysis of operational models we define the set of record-breaking events as those where
HRES-fc0 exceeds the training record, yielding 170,136 records for heat, 109,155 for cold and 338,235 for wind (Supplementary Fig.~\ref{figure::map_records_HRES_fc0_2020}).

\begin{figure}[ht!]
    \centering
    \includegraphics[width=\textwidth]{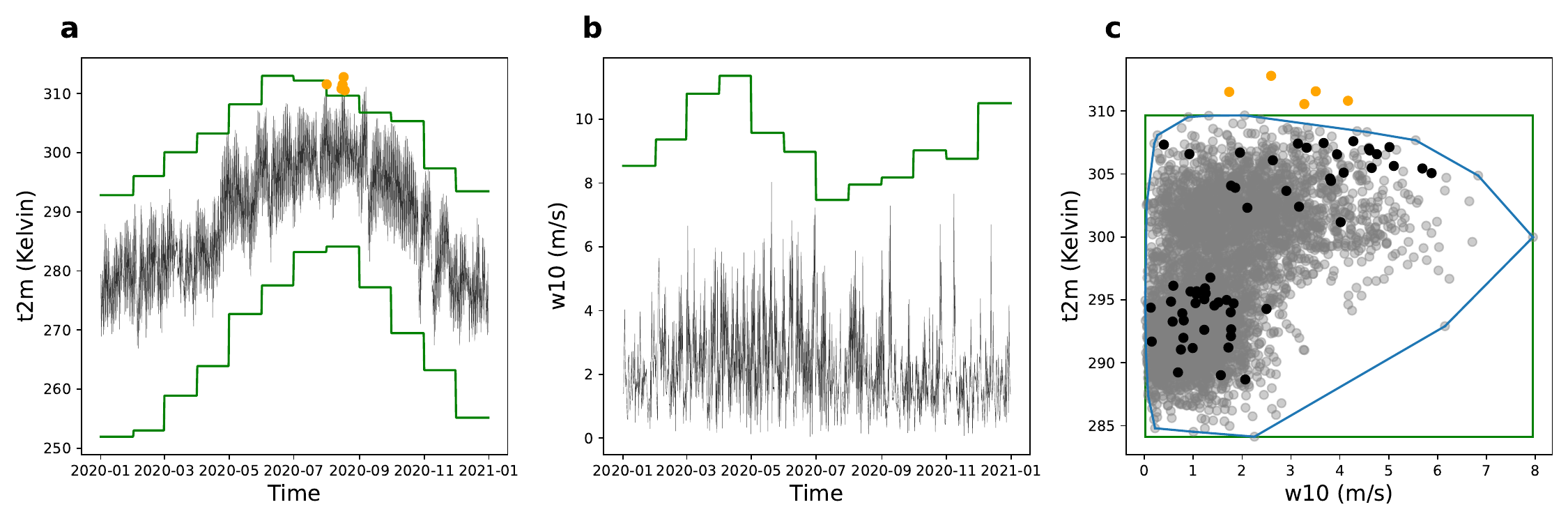}
    \caption{\textbf{Illustration of our definitions of record and extrapolation}. \textbf{a}, Daily time series of 2m temperature at the location with latitude $34.75$ and longitude $-112.25$ in 2020 (black), and monthly max/min records (in green) at this location, where orange points indicate the record-breaking events in August. \textbf{b}, Daily time series of 10m wind speed and monthly max records at the same location. \textbf{c}, Scatter plots of 2m temperature and 10m wind speed in August in the training period from 1979--2017 (in grey) and in the evaluation year 2020 (in black) at this location. The blue line represents the convex hull formed by the training data, while the green rectangle shows the max/min records in the training period. Orange points indicate the record-breaking events in the evaluation period.}
    \label{figure::intro_record_extrapolation}
\end{figure}

\subsection*{Extrapolation in AI models}

Extrapolation or out-of-distribution generalization in AI models refers to the situation where a test predictor is
far away from the distribution of the training predictors. 
In high-dimensional predictor spaces, it is not trivial to mathematically describe such points. 
One way of framing extrapolation is to require that the predictor is 
outside of the convex hull (blue line in Fig.~\ref{figure::intro_record_extrapolation}) formed by the training data.
\cite{Balestriero2021} argue that with this definition of training domain it is in fact very likely that test points need 
extrapolation. 
However, convex hulls are computationally prohibitive in high dimensions since the number of facets grows rapidly with the dimension.
Our record dataset therefore considers a stronger yet simpler definition, namely all points where at least one test variable is beyond its univariate training range. In Fig.~\ref{figure::intro_record_extrapolation} this corresponds to all test points outside of the green rectangle.
All events in the record set $R^{x,\max}$ in Equation~\eqref{R_def} satisfy this strong definition of out-of-distribution 
samples.

\subsection*{Root mean square error}
We quantify the forecast error with the root mean square error (RMSE).
For a target variable $x$ of interest (e.g., 2-meter temperature $T_{2\text{m} }$) the RMSE on a subset of location-initialization pairs for lead time $\tau$ is defined as
\begin{equation}\label{RMSE_formula}
    \text{RMSE}_I(\tau) = \sqrt{\frac{1}{\sum_{(s,t_0)\in I} \omega_{s}} \sum_{ (s,t_0) \in I} \omega_{s} (\hat x_{s,t_0}^\tau - x_{s,t_0 + \tau})^2 },
\end{equation}
where
\begin{itemize}
    \item $G_{0.25^\circ}$ is the set of locations/grid cells,
    \item $I\subseteq G_{0.25^\circ} \times T_{\text{test}}$ is the set of location-initialization pairs of interest,
    \item $\hat x_{s,t_0}^{\tau}$ is a forecast of variable $x$ with lead time $\tau$ at location $s \in G_{0.25^\circ}$ and initialization time $t_0\in T$, and $x_{s,t_0+\tau}$ is the corresponding ground truth,
    \item $\omega_{s}$ is the latitude-based weight chosen as the one used in~\cite{Lam2023}
    \begin{equation*}
        {\omega}_{s} = 
        \begin{cases}
            \cos(\theta_{\text{lat}(s)}) \sin(\theta_{0.25^{\circ}/2}), &\text{ if } |\theta_{\text{lat}(s)}| < \pi/2, \\
            \sin^2(\theta_{0.25^{\circ}/4}), &\text{ if } |\theta_{\text{lat}(s)}| = \pi/2,
        \end{cases} 
    \end{equation*}
    with $\theta_{a}$ as the radian associated with degree $a$. 
\end{itemize}

Our definition of RMSE is more general than the conventional one~\citep{Rasp2024} in the sense that we allow to focus on a subset $I$ of location-initialization pairs $(s,t_0)$.
If we set $I$ as the product of the set of all grid cells over the globe and all time points in $T_{\text{test}}$, 
we recover the traditional (latitude-weighted) RMSE on all test locations and initialization times. 

In the computation of RMSE on record-breaking events such as shown in Fig.~\ref{figure::RMSE_t2m_w10_2020}, we choose the set $I$ in Equation~\eqref{RMSE_formula} in the following way.
Recall the set $R^{x,\max}\subseteq G_{0.25^\circ} \times T_{\text{test}}$ of location-time pairs of all records in a given time period (e.g., the year 2020).
We choose all location-initialization pairs such that the target of the forecasting with lead time $\tau$ corresponds to a record, i.e.,
\begin{align}\label{I_tau}
    I_\tau = \left\{ (s,t_0) \in G_{0.25^\circ} \times T_{\text{test}}: (s, t_0 + \tau) \in R^{x,\max}  \right\}.
\end{align} 
The corresponding $\text{RMSE}_{I}(\tau)$ is the error of a model made in forecasting records with lead time $\tau$.

To construct a confidence interval for the $\text{RMSE}_I(\tau)$ in Equation~\eqref{RMSE_formula}, we assume that the central limit theorem holds for the weighted square errors, i.e.,
\begin{equation*}
    \sqrt{ |I| } \Bigg[ \frac{1}{ \sum_{(s,t_0)\in I} \omega_{s} } \sum_{ (s,t_0) \in I} \omega_{s}(\hat x_{s,t_0}^\tau - x_{s,t_0 + \tau})^2 - \mu_{\tau} \Bigg]
    \Dconvergence \mathcal{N}(0, \sigma_{\tau}^2), \quad |I| \rightarrow \infty,
\end{equation*}
where $\mu_{\tau}$ denotes the true mean of weighted square error and $\sigma_{\tau}^2$ its asymptotic variance, and $\Dconvergence$ means convergence in distribution.
Then by the Delta method, we obtain the asymptotic distribution of $\text{RMSE}_I(\tau)$
\begin{equation*}
    \sqrt{ |I| } \big(\text{RMSE}_I(\tau) - \sqrt{\mu_{\tau}} \big) \Dconvergence \mathcal{N}\big(0, \sigma_{\tau}^2/(4\mu_{\tau}) \big).
\end{equation*}
Hence an approximate $\alpha$-level confidence interval for $\text{RMSE}_I(\tau)$ can be constructed as $[\mu_{\tau} - q_{(1+\alpha)/2}\sigma_{\tau}/\sqrt{4\mu_{\tau} |I|}, \mu_{\tau} + q_{(1+\alpha)/2}\sigma_{\tau}/\sqrt{4\mu_{\tau} |I|}]$, where $q_{(1+\alpha)/2}$ is the $(1+\alpha)/2$-quantile of a standard normal distribution.
Alternatively, bootstrap can be used to construct the confidence bands~\citep{Bodnar2025}. We tried the non-parametric bootstrap with 1000 resampling, which yielded similar confidence bands to the normal ones. For the sake of computational feasibility, we use the above normal confidence levels throughout the paper.

\subsection*{Forecast bias}
To complement RMSE and investigate whether a forecasting model under- or overpredicts the ground truth, we consider the latitude-weighted forecast bias
\begin{equation*}
    \text{FB}_I(\tau) = \frac{1}{ \sum_{(s,t_0)\in I} {\omega_{s}}} \sum_{ (s,t_0) \in I} \omega_{s} (\hat x_{s,t_0}^{\tau} - x_{s, t_0+\tau}).
\end{equation*}
where the notation is the same as in Equation~\eqref{RMSE_formula}. Confidence intervals for the forecast bias are computed in the same way as for
RMSE based on asymptotic normality.

\subsection*{Precision and recall curves}

For early warning systems it is crucial that a weather forecasting model is able to predict the occurrence of an extreme event accurately. We therefore consider record-breaking event forecasting as a binary classification problem by assessing whether a forecasting model can predict the exceedance of a variable over its previous record in the sense of~\eqref{record_def}. 
Since this classification problem is strongly imbalanced, similar to previous studies~\citep{Lam2023}, we use precision-recall curves that are well-suited for such cases since they account for both false positives and false negatives.

For a set $I\subseteq G_{0.25^\circ}\times T_{\text{test}}$ of location-initialization pairs of interest, we compute the precision and recall
for variable $x$ at lead time $\tau$ as
(we set $r_{s,m} = r^{x, \max}_{s,m}$ to simplify notation)
\begin{align*}
    \text{Precision}_I(\tau) &= \frac{ \sum_{(s,t_0)\in I} \mathbf 1 \{ \hat x_{s,t_0}^\tau > r_{s,m(t_0+\tau)}\} \mathbf 1 \{ x_{s,t_0 + \tau} > r_{s,m(t_0+\tau)} \}} 
    { \sum_{(s,t_0)\in I} \mathbf 1 \{ \hat x_{s,t_0}^\tau > r_{s,m(t_0+\tau)}\} }, 
\end{align*}
and 
\begin{align*}
    \text{Recall}_I(\tau) &= \frac{ \sum_{(s,t_0)\in I} \mathbf 1 \{ \hat x_{s,t_0}^\tau > r_{s,m(t_0+\tau)}\} \mathbf 1 \{ x_{s,t_0 + \tau} > r_{s,m(t_0+\tau)} \}} 
    { \sum_{(s,t_0)\in I} \mathbf 1 \{ x_{s,t_0 + \tau} > r_{s,m(t_0+\tau)}\} }.
\end{align*}
where $\hat x_{s,t_0}^\tau$ denotes a forecast of variable $x$ at location $s$ initialized at time $t_0$ with lead time $\tau$, and $x_{s,t_0 + \tau}$ is the corresponding ground truth. As above $m(t_0+\tau)$ is the month corresponding to time $t_0 + \tau$. 

In order to produce a precision-recall curve from a deterministic forecast, following~\cite{Lam2023}, we introduce a common ``gain" parameter to define scaled forecasts by 
\begin{align}\label{scaled_forecast}
    \text{scaled forecast = forecast + gain × forecast std.~deviation}.
\end{align}
Using these scaled forecasts in the precision and recall formulae instead of only $\hat x_{s,t_0}^\tau$ and varying the gain parameter in a suitable range ($[-1.5, 1.5]$ in our case) yields a precision-recall curve.
The scaling allows the study of different trade-offs between false positives and false negatives, and using a common gain parameter enables averaging over all spatial locations $s\in G_{0.25^\circ}$.
Our parameterization of the scaled forecasts is slightly different from the one in~\cite{Lam2023}, but it is theoretically more justified.
Indeed, for a probabilistic forecast from a location-scale family, formula~\eqref{scaled_forecast} corresponds to choosing the same quantile of the forecast distribution at all locations.

For each variable $x$, each location $s\in G_{0.25^\circ}$, each month $m$ and each lead time $\tau$ we estimate the forecast standard deviations in~\eqref{scaled_forecast}
from forecasts in the year 2020 for the different models.
We assume that this standard deviation is constant for time points in the same month so that we have enough data for the estimation.

\subsection*{Correlation between record forecasts}

In order to compute the correlation between the 
different model forecasts and ground truths, we 
define suitable functions indicating whether the corresponding
record is exceeded. Fix a lead time $\tau$ and, for instance, consider the variable $x$ with max-records abbreviated by $r_{s,m} = r_{s,m}^{x,\max}$.
For a forecasts $\hat x_{s,t}^\tau$ from some model define
for each time point $t_0\in T_{\text{test}}$ the indicator $\mathbf{1}\{\hat x_{s,t_0}^\tau > r_{s,m(t_0+\tau)} \}$ that takes
value $1$ if $\hat x_{s,t_0}^\tau$ exceeds the record $r_{s,m(t_0+\tau)}$
and $0$ otherwise. We can now compute the correlation between 
these indicators, indexed by all $t_0+\tau\in T_{\text{test}}$ and $s\in G_{0.25^\circ}$, for two different forecast models.
Similarly, for a ground truth (either ERA5 or HRES-fc0) we define
for each time point $t_0+\tau\in T_{\text{test}}$ the indicator $\mathbf{1}\{x_{s,t_0+\tau} >r_{s,m(t_0+\tau)} \}$.
We then compute correlations of these ground truths with 
the forecast indicators, and between forecast indicators from different models. The resulting correlation matrix
is shown for the ERA5-trained AI models
in Fig.~\ref{figure::precision_recall_GraphCast}g--i, and for the operational AI models in Supplementary Fig.~\ref{figure::precision_recall_correlation_GraphCast_oper}g--i.

\subsection*{Forecaster's dilemma}

In the theory of forecast evaluation, the forecaster's dilemma \citep{Lerch2017} shows that computing an evaluation score only on a subset of observations can incentivize sub-optimal forecasts.
Such conditioning appears in the RMSE defined in~\eqref{RMSE_formula} if the set $I$ depends on the observations, as, for instance, in the case of record-breaking events defined in~\eqref{I_tau}.
This metric should therefore not be used as the sole evaluation criterion, but rather in combination with others.
We therefore also report the overall RMSE on all events in Fig.~\ref{figure::RMSE_t2m_w10_2020}, Supplementary Figs.~\ref{figure::RMSE_t2m_w10_2020_oper} and~\ref{figure::RMSE_t2m_w10_2018}, which show that all methods yield errors on a comparable scale and do not appear to artificially hedge forecasts of extreme events.
In addition, we consider different evaluation criteria
such as precision-recall curves that take into account 
both false positives and false negatives (Fig.~\ref{figure::precision_recall_GraphCast}d--f and Supplementary Fig.~\ref{figure::precision_recall_correlation_GraphCast_oper}d--f).

Computing the RMSE on a subset of extreme observations 
is common in the literature of AI weather forecasts~\citep{Lam2023,Olivetti2024,Bodnar2025}.
Another approach that avoids the forecaster's dilemma
completely is to condition on the forecasts instead
of the observations~\citep{Holzmann2014}.
We follow this approach to compare the operational
version of GraphCast with HRES. We choose as the 
set of record-breaking events in~\eqref{R_def}
all location-initialization pairs such that forecasts
with lead time $\tau$
from both GraphCast operational and HRES exceed
the training record. For max-records of variable $x$, for instance,
this yields an index set
\begin{align*}
    I_\tau = \left\{ (s,t_0) \in G_{0.25^\circ} \times T_{\text{test}}: \hat x_{s,t_0}^{\text{HRES},\tau} > r_{s,m(t_0+\tau)}, \hat x_{s,t_0}^{\text{GraphCast},\tau} > r_{s,m(t_0+\tau)}  \right\},
\end{align*}
to be used in the RMSE in~\eqref{RMSE_formula}.
The results (Supplementary Fig.~\ref{figure::RMSE_forecast_bias_2020_oper_cond_HRES_GC})
look qualitatively similar to those from
conditioning on the observations, except that we have 
a much smaller set of events that are jointly forecasted to be
record-breaking by both models compared to the original record
dataset.

\printbibliography
\end{refsection}




\let\cleardoublepage\clearpage
\appendix

\newpage
\renewcommand{\contentsname}{Supplementary materials}
\let\cleardoublepage\clearpage
\tableofcontents

\newpage
\section{Supplementary figures for the evaluation of non-operational forecasts in 2020}

\begin{figure}[H]  
\renewcommand\figurename{Supplementary Fig.} 
    \centering
    \includegraphics[width=\textwidth]{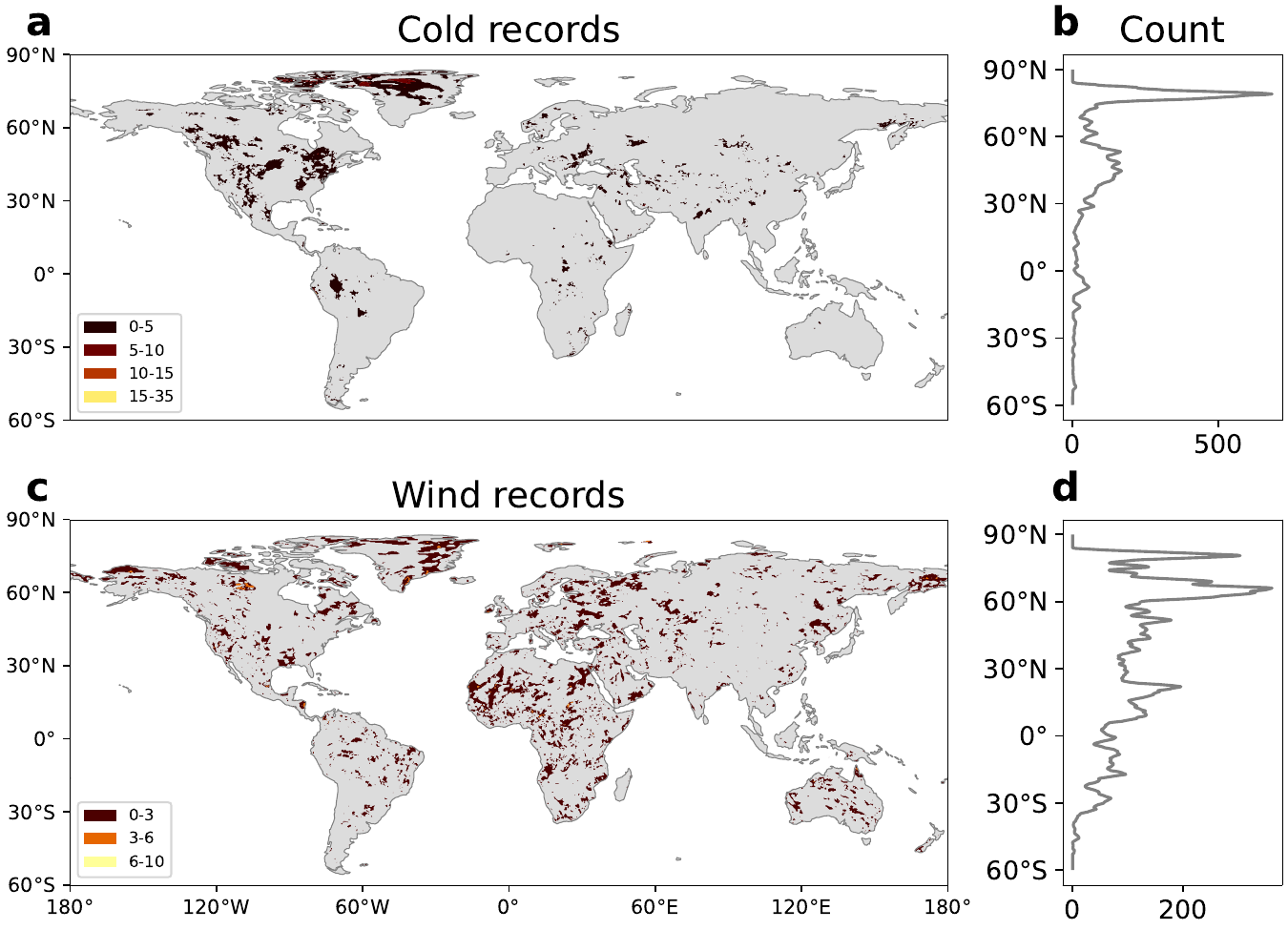}
    \caption{\textbf{Number of record-breaking events over land (excluding the Antarctic region) in 2020 in ERA5}. \textbf{a} and \textbf{c}, Number of cold and wind records. \textbf{b} and \textbf{d}, Number of cold and wind records per latitude.}
    \label{figure::map_records_ERA5_2020}
\end{figure}

\begin{figure}[H] 
\renewcommand\figurename{Supplementary Fig.} 
    \centering
    \includegraphics[width=0.95\textwidth]{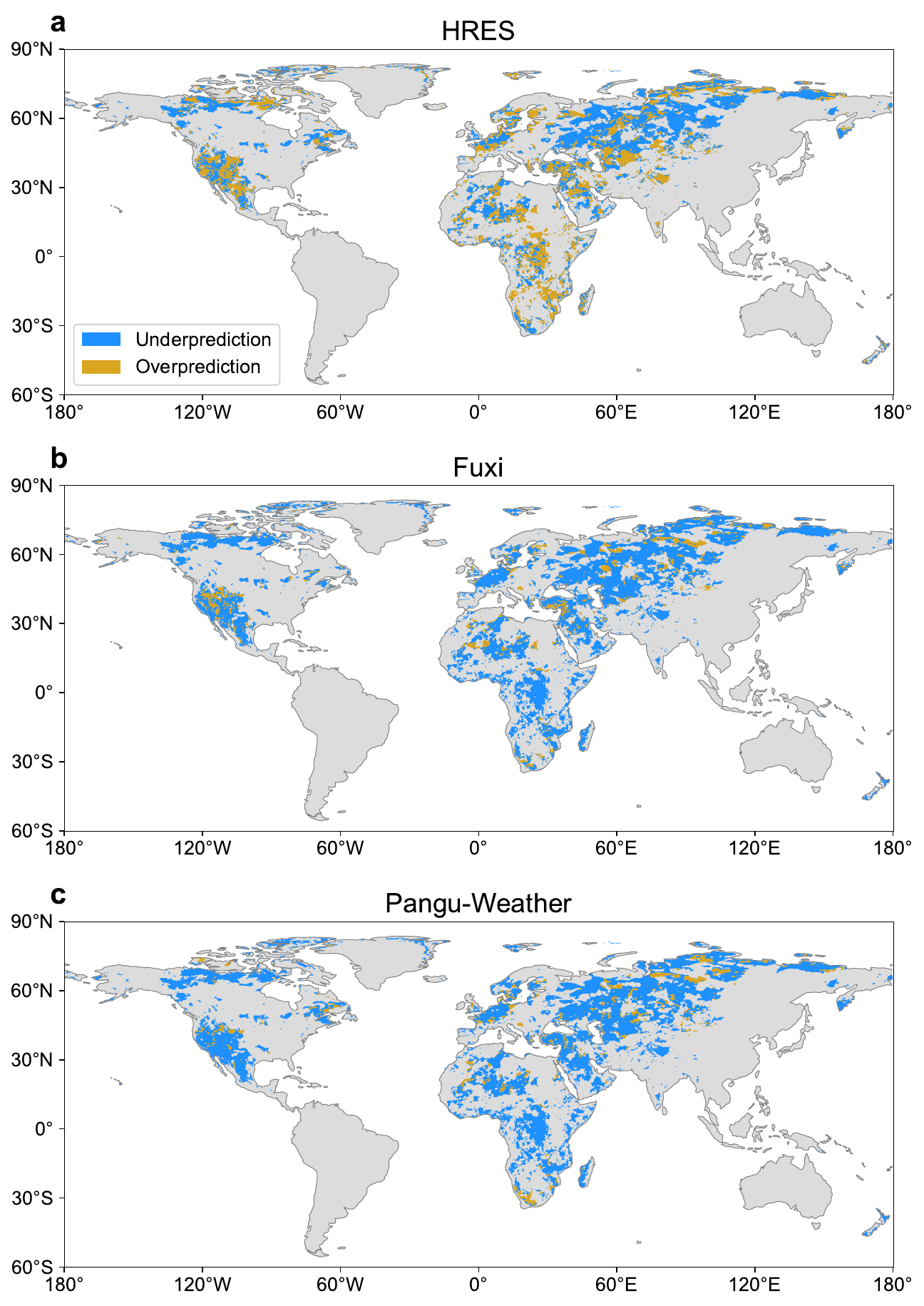}
    \caption{\textbf{Forecast bias of the maximum heat records for different models in 2020.}. \textbf{a}, Forecast bias for the numerical model HRES. \textbf{b}, Forecast bias for AI model Fuxi. \textbf{c}, Forecast bias for AI model Pangu-Weather.
    }
    \label{figure::map_under_over_prediction_2020_non_oper}
\end{figure}

\begin{figure}[H] 
\renewcommand\figurename{Supplementary Fig.} 
    \centering
    \includegraphics[width=0.95\textwidth]{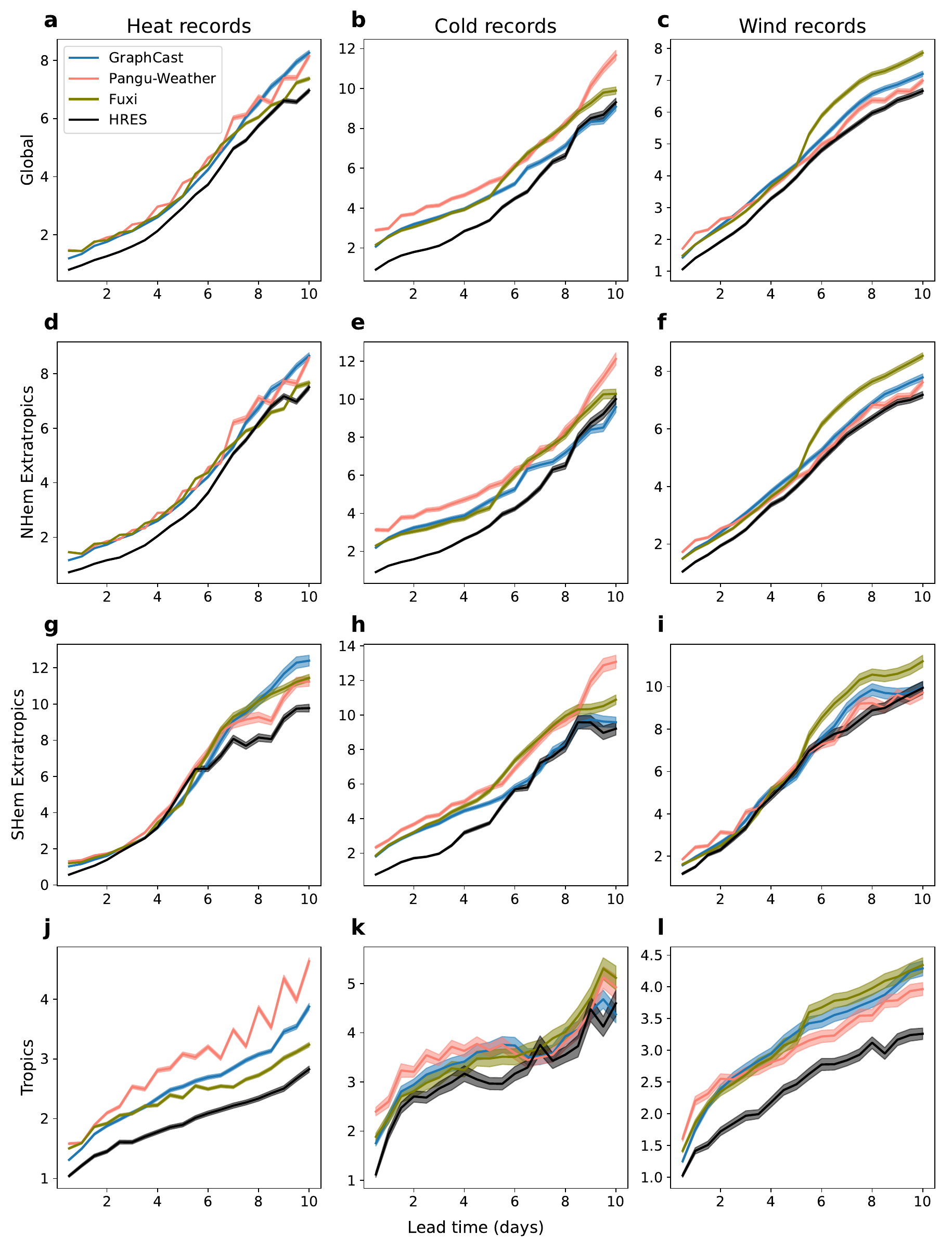}
    \caption{\textbf{Regional RMSE of 2m temperature and 10m wind speed over land in 2020}.
    \textbf{a}--\textbf{c}, RMSE for the whole globe.
    \textbf{d}--\textbf{f}, RMSE for Northern Hemisphere Extratropics (NHem Extratropics, i.e., grid cells with latitude in $[20, 90]$).
    \textbf{g}--\textbf{i}, RMSE for Southern Hemisphere Extratropics (SHem Extratropics, i.e., grid cells with latitude in $[-90, -20]$).
    \textbf{j}--\textbf{l}, RMSE for Tropics (grid cells with latitude in $[-20, 20]$).
    The transparent shaded areas indicate $95\%$ confidence bands. 
    }
    \label{figure::RMSE_regional_2020}
\end{figure}

\begin{figure}[H] 
\renewcommand\figurename{Supplementary Fig.} 
    \centering
    \includegraphics[width=0.95\textwidth]{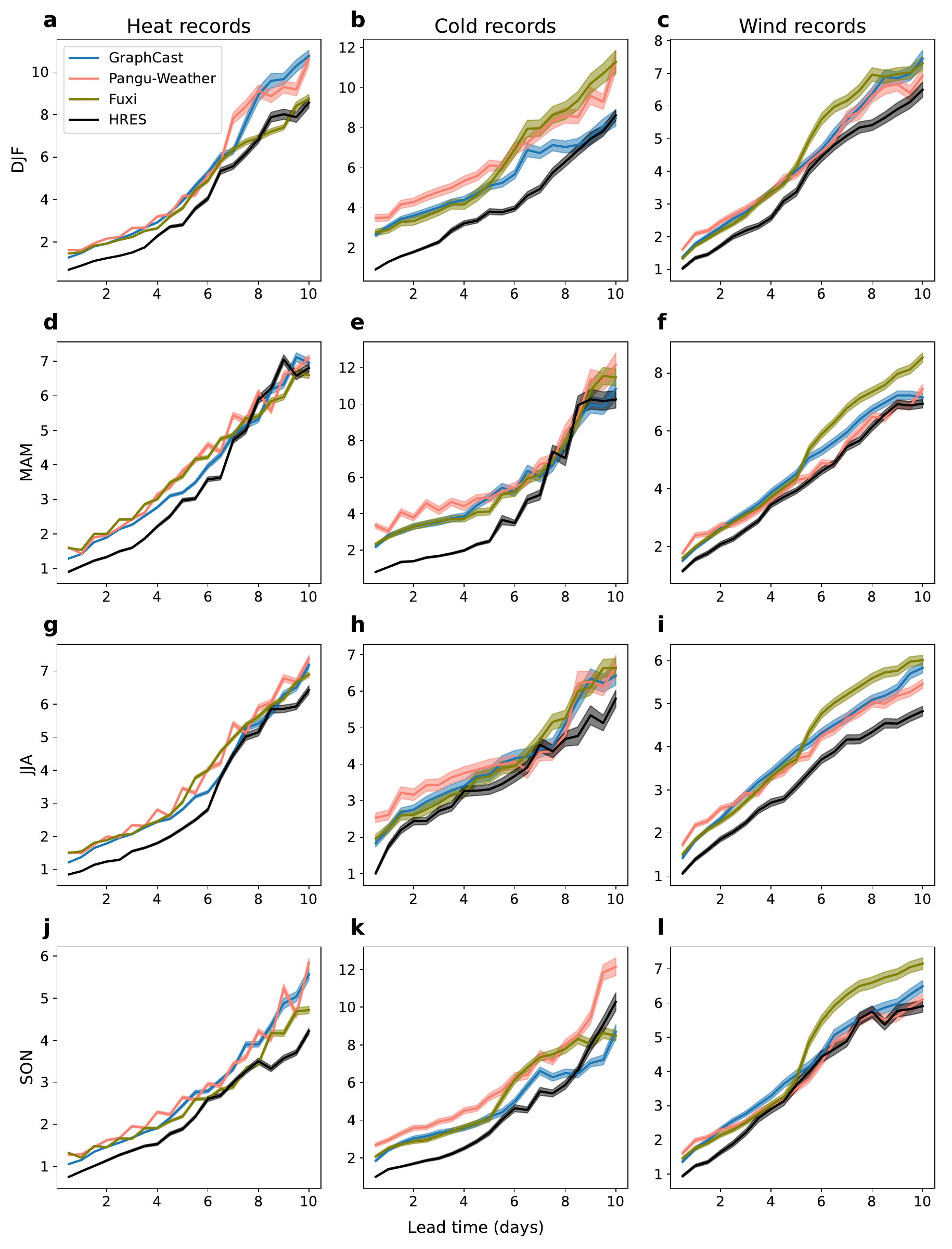}
    \caption{\textbf{Seasonal RMSE of 2m temperature and 10m wind speed over land (excluding the Antarctic region) in 2020}.
    \textbf{a}--\textbf{c}, RMSE for the months December -- February.
    \textbf{d}--\textbf{f}, RMSE for the months March -- May.
    \textbf{g}--\textbf{i}, RMSE for the months June -- August.
    \textbf{j}--\textbf{l}, RMSE for the months September -- November. 
    }
    \label{figure::RMSE_seasonal_2020}
\end{figure}

\begin{figure}[H] 
\renewcommand\figurename{Supplementary Fig.} 
    \centering
    \includegraphics[width=\textwidth]{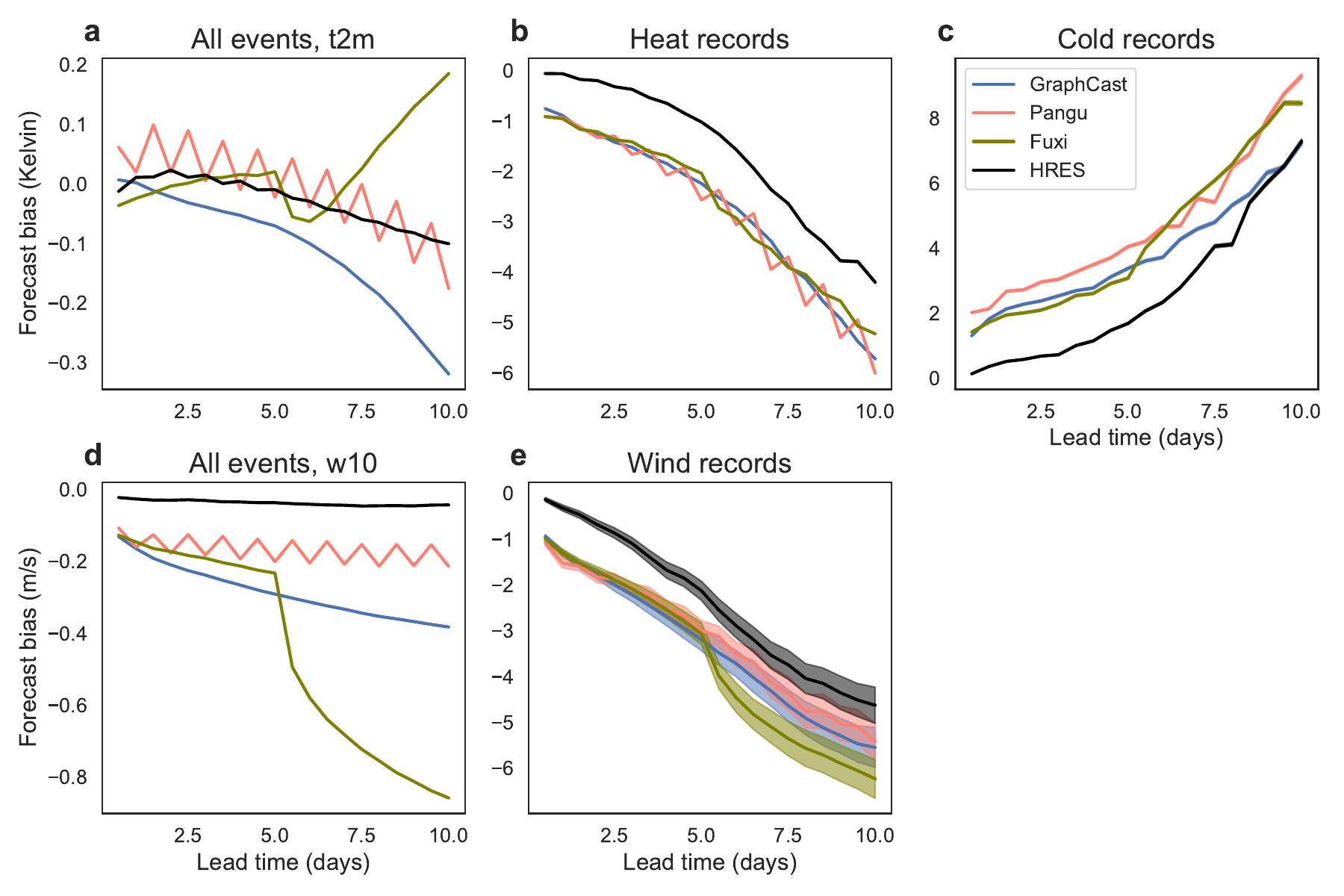}
    \caption{\textbf{Forecast bias for all and record-breaking events in 2020}. Forecast bias of 2m temperature and 10m wind speed over land (excluding the Antarctic region) of HRES, GraphCast, Pangu-Weather, and Fuxi for all events (\textbf{a}, \textbf{d}) and only record-breaking events (\textbf{b}, \textbf{c}, \textbf{e}) in 2020. The transparent shaded areas indicate $95\%$ confidence bands.
    }
    \label{figure::FB_all_records_2020}
\end{figure}

\begin{figure}[H] 
\renewcommand\figurename{Supplementary Fig.} 
    \centering
    \includegraphics[width=\textwidth]{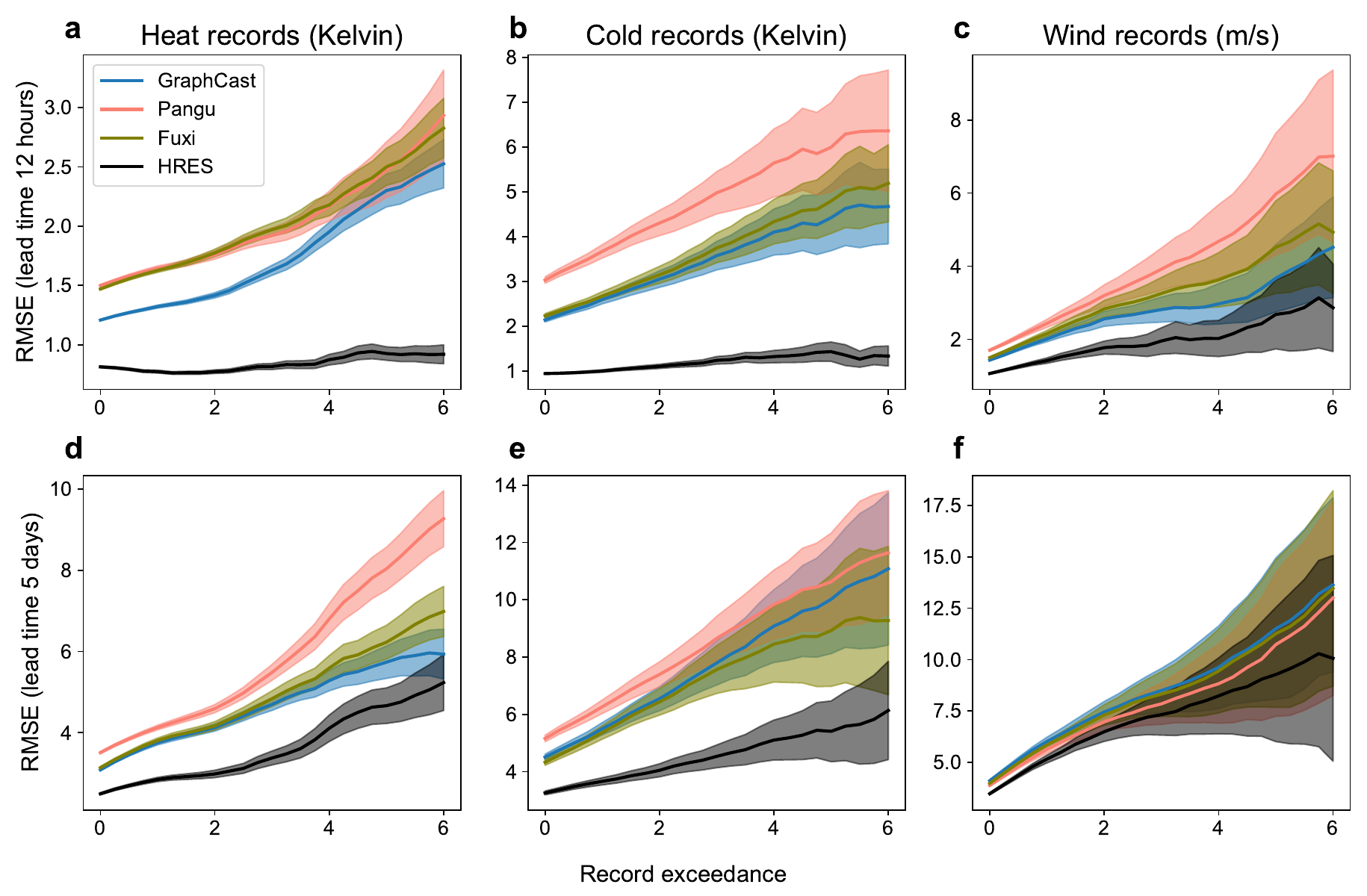}
    \caption{\textbf{RMSE against record exceedance for different lead times in 2020.} RMSE of 2m temperature and 10m wind speed for lead time 12 hours (\textbf{a}--\textbf{c}) and 5 days (\textbf{d}--\textbf{f}) over land (excluding the Antarctic region) in 2020, for events that exceed the record by at least a certain margin (in Kelvin or m/s). The transparent shaded areas indicate $95\%$ confidence bands.
    }
    \label{figure::RMSE_record_exceedance_2020}
\end{figure}

\begin{figure}[H] 
\renewcommand\figurename{Supplementary Fig.} 
    \centering
    \includegraphics[width=\textwidth]{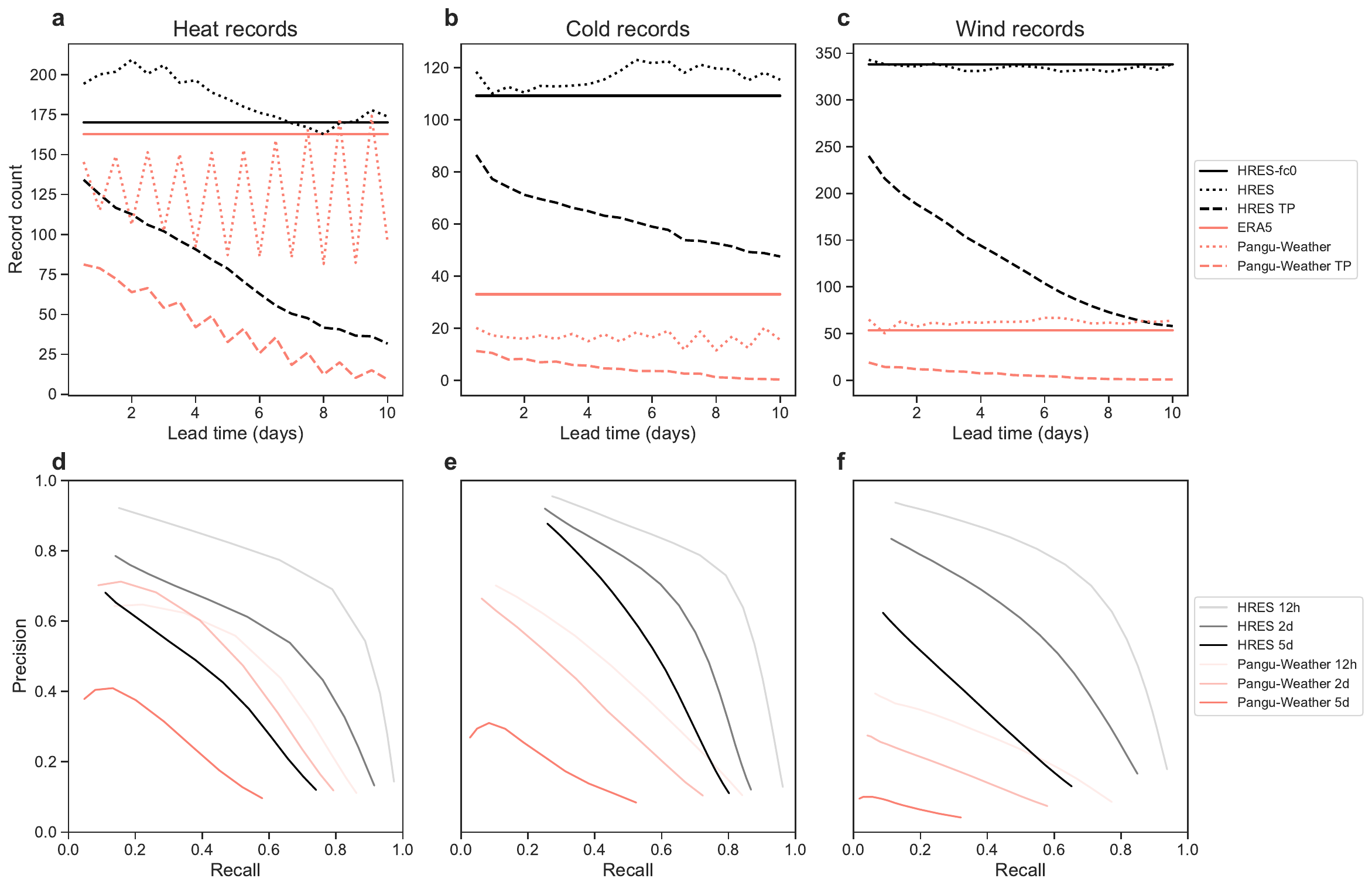}
    \caption{\textbf{Pangu-Weather forecast of occurrence of record-breaking events over land (excluding the Antarctic region) in 2020}. $\textbf{a}$-$\textbf{c}$, Counts (in thousands) of heat, cold, and wind records in the ground truth ERA5 and HRES-fc0, and Pangu-Weather and HRES forecast data, as well as counts of their true positives (TP). $\textbf{d}$-$\textbf{f}$, Precision and recall curve of Pangu-Weather and HRES forecasts when using the record data as the threshold. }
    \label{figure::precision_recall_Pangu_2020_non_oper}
\end{figure}

\begin{figure}[H] 
\renewcommand\figurename{Supplementary Fig.} 
    \centering
    \includegraphics[width=\textwidth]{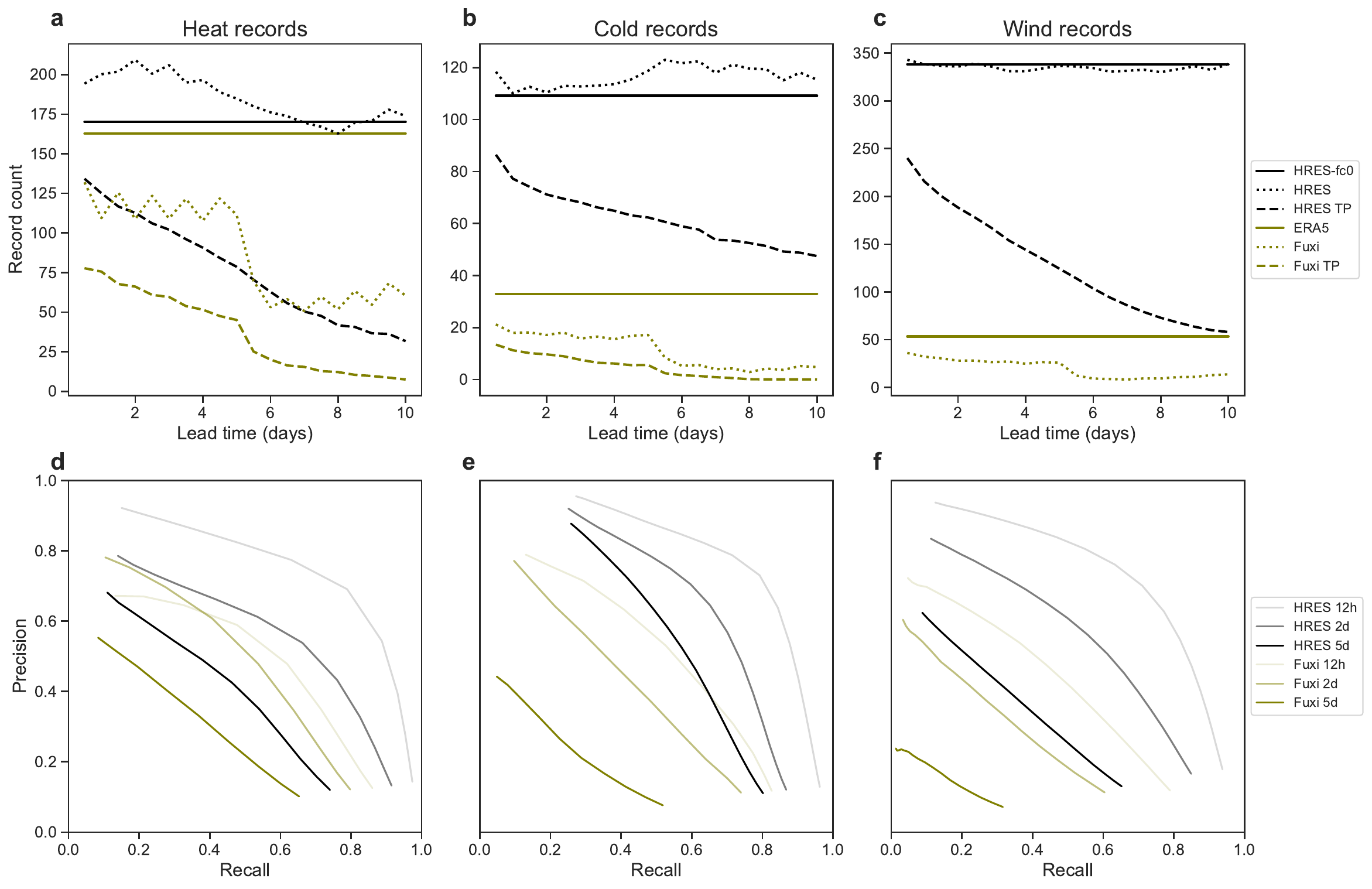}
    \caption{\textbf{Fuxi forecast of occurrence of record-breaking events over land (excluding the Antarctic region) in 2020}. $\textbf{a}$-$\textbf{c}$, Counts (in thousands) of heat, cold, and wind records in the ground truth ERA5 and HRES-fc0, and Fuxi and HRES forecast data, as well as counts of their true positives (TP). $\textbf{d}$-$\textbf{f}$, Precision and recall curve of Fuxi and HRES forecasts when using the record data as the threshold.}
    \label{figure::precision_recall_Fuxi_2020_non_oper}
\end{figure}

\section{Supplementary figures for the evaluation of operational forecasts in 2020}

\begin{figure}[H] 
\renewcommand\figurename{Supplementary Fig.} 
    \centering
    \includegraphics[width=\textwidth]{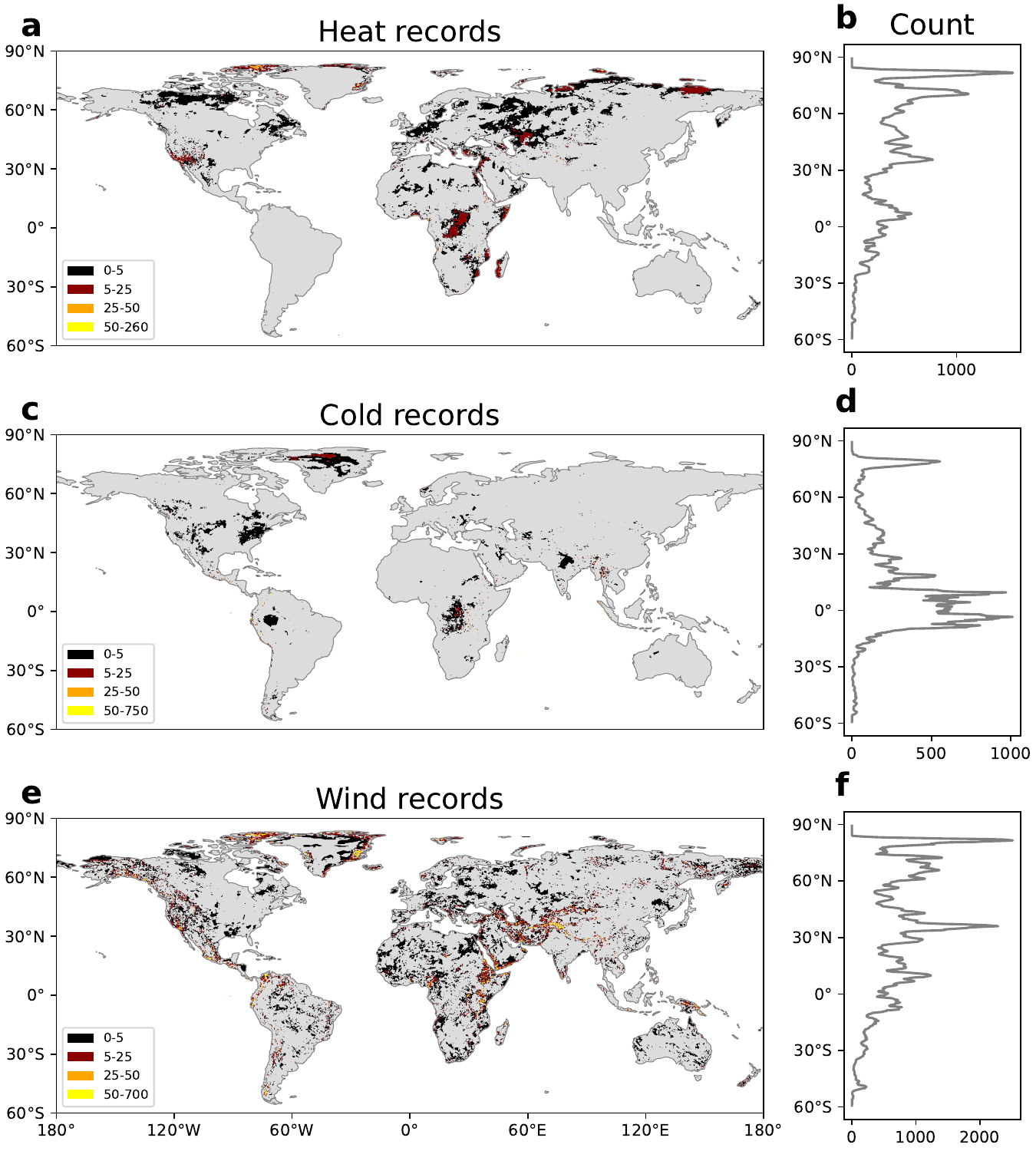}
    \caption{\textbf{Number of records over land (excluding the Antarctic region) in 2020 in HRES-fc0}. \textbf{a}, \textbf{c}, \textbf{e}, Number of heat, cold, and wind records. \textbf{b}, \textbf{d}, \textbf{f}, Number of heat, cold, and wind records per latitude.}
    \label{figure::map_records_HRES_fc0_2020}
\end{figure}

\begin{figure}[H] 
\renewcommand\figurename{Supplementary Fig.} 
    \centering
    \includegraphics[width=\textwidth]{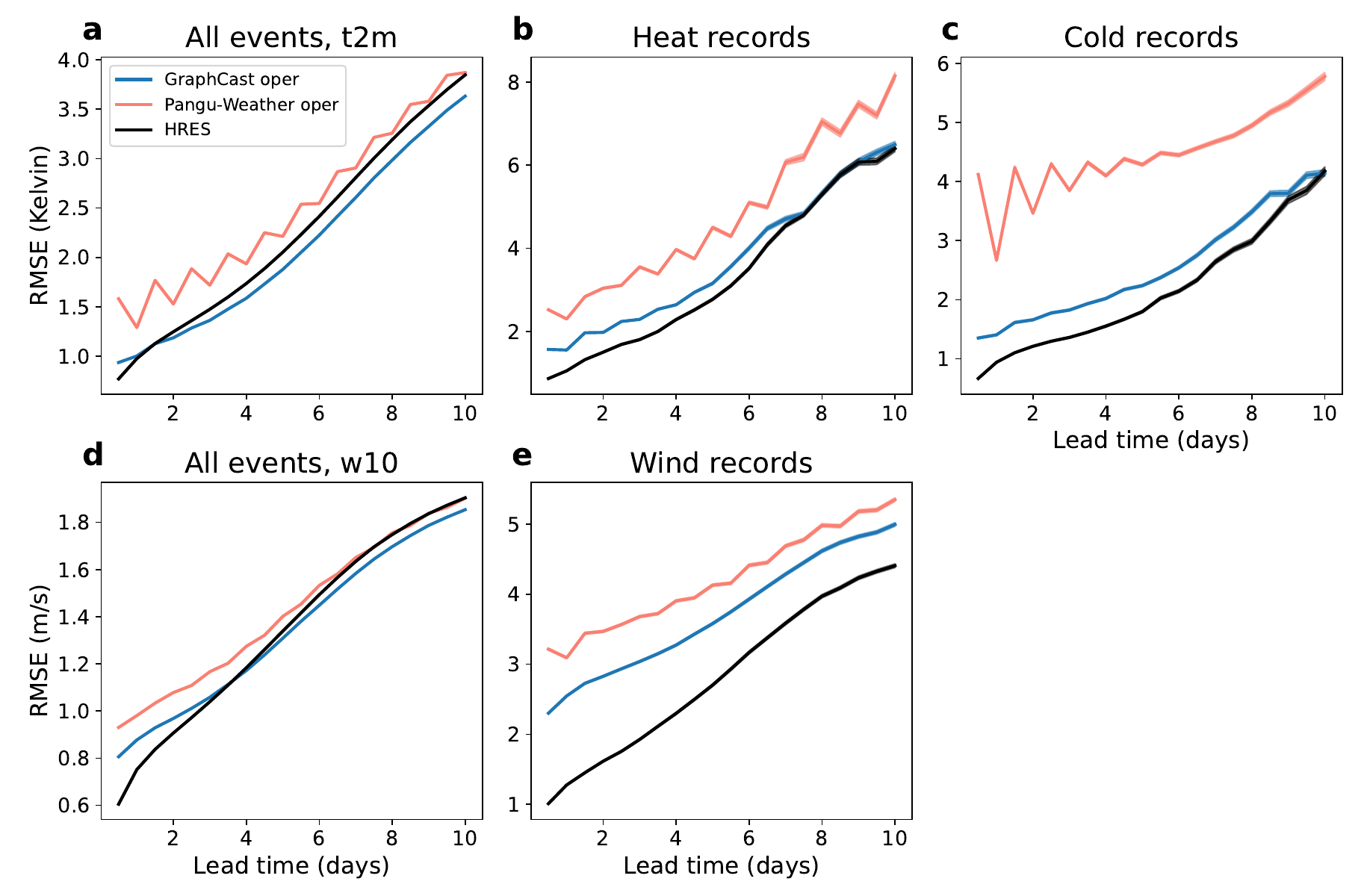}
    \caption{\textbf{Model performance of operational forecasts on all events and record-breaking events in 2020.}  RMSE of 2m temperature and 10m wind speed over land (excluding the Antarctic region) of HRES, Pangu-Weather operational, and GraphCast operational for all events (\textbf{a}, \textbf{d}) and only record-breaking events (\textbf{b}, \textbf{c}, \textbf{e}) in 2020. The transparent shaded areas indicate $95\%$ confidence bands.
    }
    \label{figure::RMSE_t2m_w10_2020_oper}
\end{figure}

\begin{figure}[H] 
\renewcommand\figurename{Supplementary Fig.} 
    \centering
    \includegraphics[width=\textwidth]{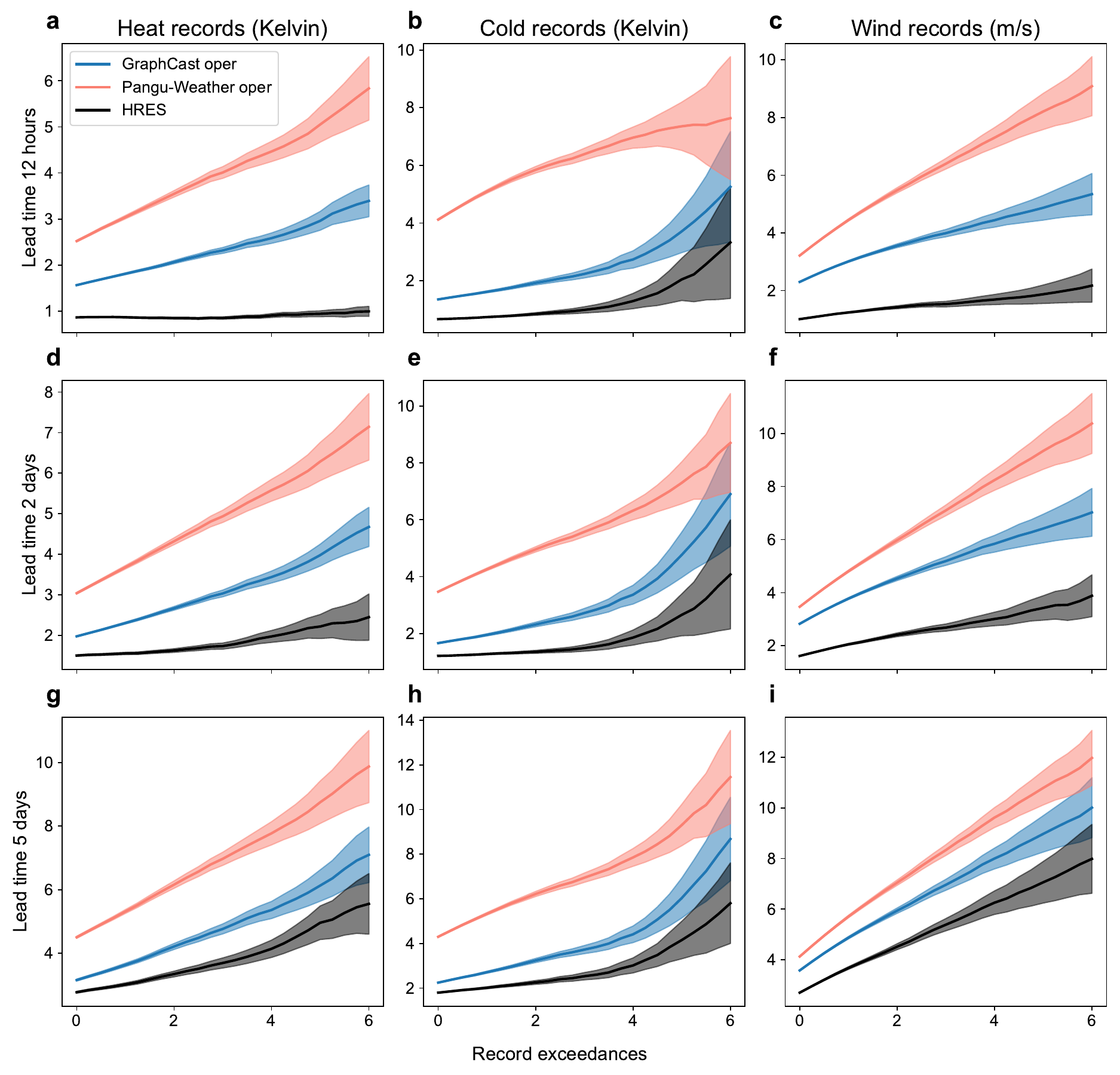}
    \caption{\textbf{RMSE of operational forecasts against record exceedances for different lead times in 2020}. RMSE of 2m temperature and 10m wind speed over land (excluding the Antarctic region) of HRES, Pangu-Weather operational, and GraphCast operational in 2020, for events that exceed the record at least by a certain margin, for lead time 12 hours (\textbf{a}--\textbf{c}), 2 days (\textbf{d}--\textbf{f}), 5 days (\textbf{g}--\textbf{i}). 
    The transparent shaded areas indicate $95\%$ confidence bands.
    }
    \label{figure::RMSE_record_exceedance_operational}
\end{figure}

\begin{figure}[H]
\renewcommand\figurename{Supplementary Fig.} 
    \centering
    \includegraphics[width=\textwidth]{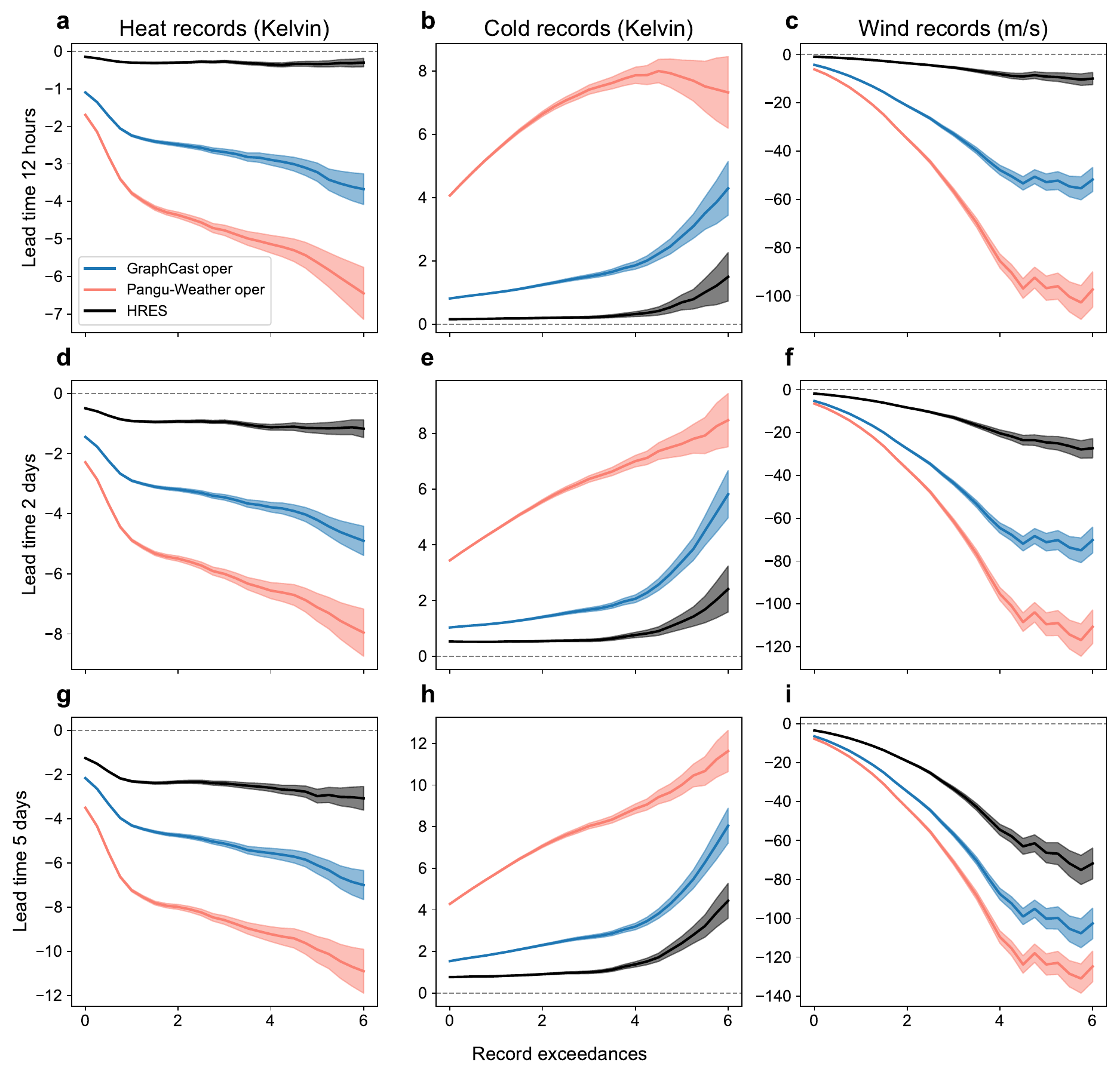}
    \caption{\textbf{Forecast bias of operational forecasts against record exceedances for different lead times in 2020}. Forecast bias of 2m temperature and 10m wind speed over land (excluding the Antarctic region) of HRES, Pangu-Weather operational, and GraphCast operational in 2020, for events that exceed the record at least by a certain margin, for lead time 12 hours (\textbf{a}--\textbf{c}), 2 days (\textbf{d}--\textbf{f}), forecast bias for lead time . \textbf{g}--\textbf{i}, forecast bias for lead time 5 days. 
    The transparent shaded areas indicate $95\%$ confidence bands.
    }
    \label{figure::FB_record_exceedance_operational}
\end{figure}

\begin{figure}[H]
\renewcommand\figurename{Supplementary Fig.} 
    \centering
    \includegraphics[width=\textwidth]{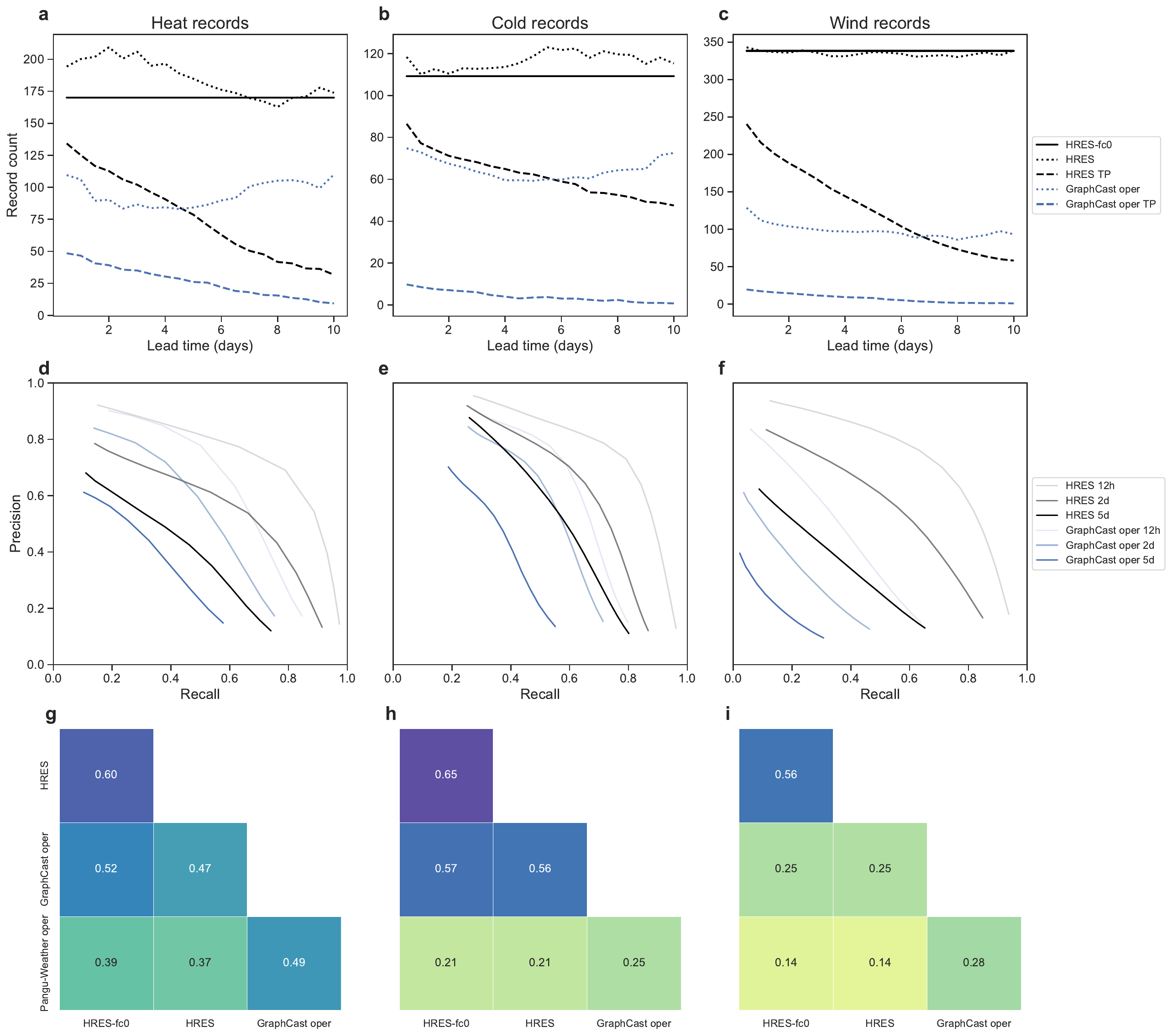}
    \caption{\textbf{Operational forecasts of the occurrence of record-breaking events}. \textbf{a}--\textbf{c}, counts (in thousands) of heat, cold, and wind records over land (excluding the Antarctic region) in 2020 in the ground truth HRES-fc0 data, and GraphCast operational and HRES forecast data, as well as counts of their true positives (TP). \textbf{d}--\textbf{f}, Precision and recall curve of GraphCast operational and HRES forecasts when the record data are used as the threshold. 
    \textbf{g}--\textbf{i}, Correlations between the indicator functions of whether the ground truth or 2-day forecast data exceed the record. 
    }
    \label{figure::precision_recall_correlation_GraphCast_oper}
\end{figure}

\begin{figure}[H]
\renewcommand\figurename{Supplementary Fig.} 
    \centering
    \includegraphics[width=\textwidth]{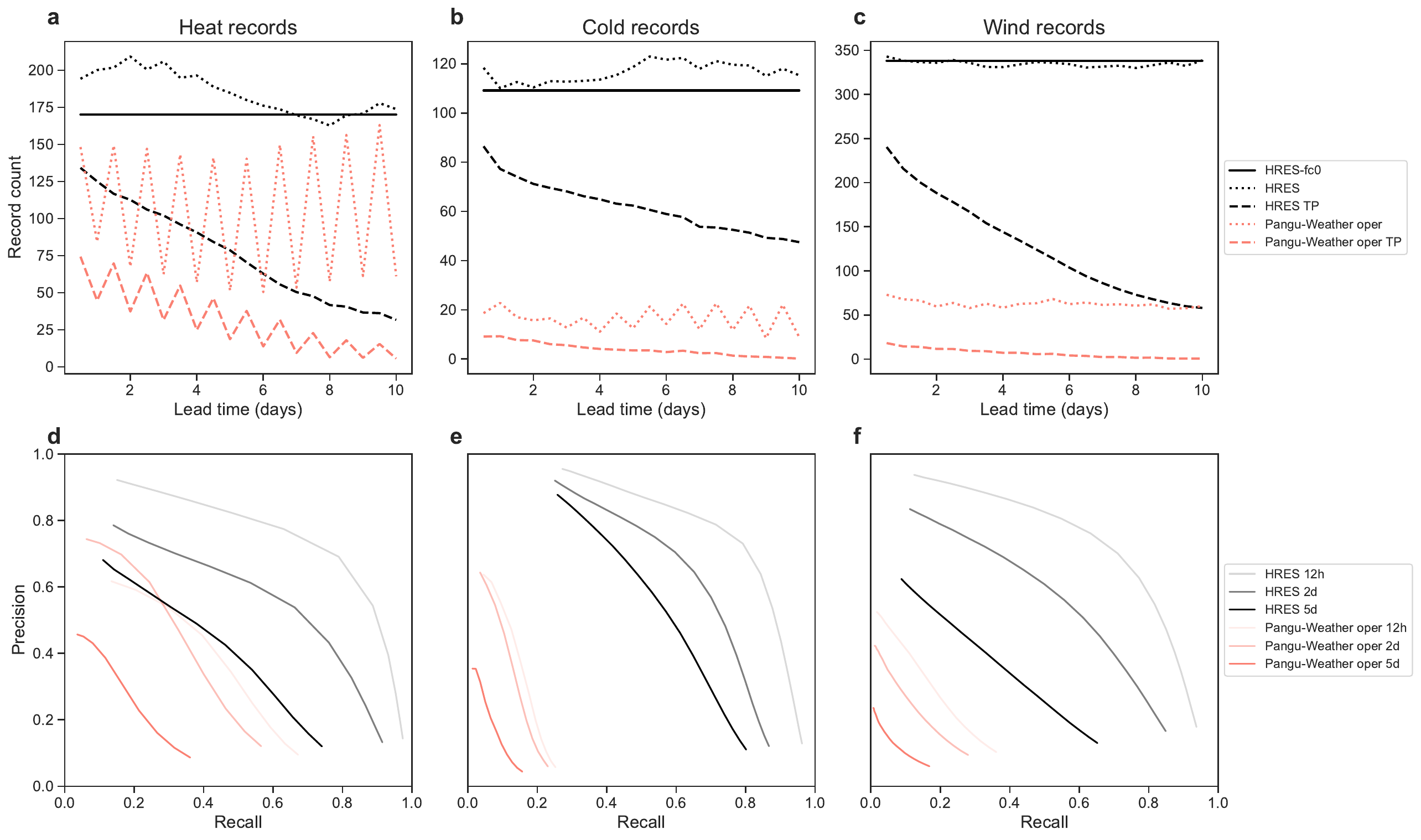}
    \caption{\textbf{Pangu-Weather operational forecast of the occurrence of record-breaking events over land (excluding the Antarctic region) in 2020}. \textbf{a}--\textbf{c}, Counts (in thousands) of heat, cold, and wind records in the ground truth HRES-fc0 data, and Pangu operational and HRES forecast data, as well as counts of their true positives (TP). \textbf{d}--\textbf{f}, Precision and recall curve of Pangu operational and HRES forecasts when the record data are used as the threshold. 
    \textbf{g}--\textbf{i}, Correlations between the indicator functions of whether the ground truth or 2-day forecast data exceed the record. 
    }
    \label{figure::precision_recall_correlation_Pangu_oper}
\end{figure}

\begin{figure}[H]
\renewcommand\figurename{Supplementary Fig.} 
    \centering
    \includegraphics[width=0.95\textwidth]{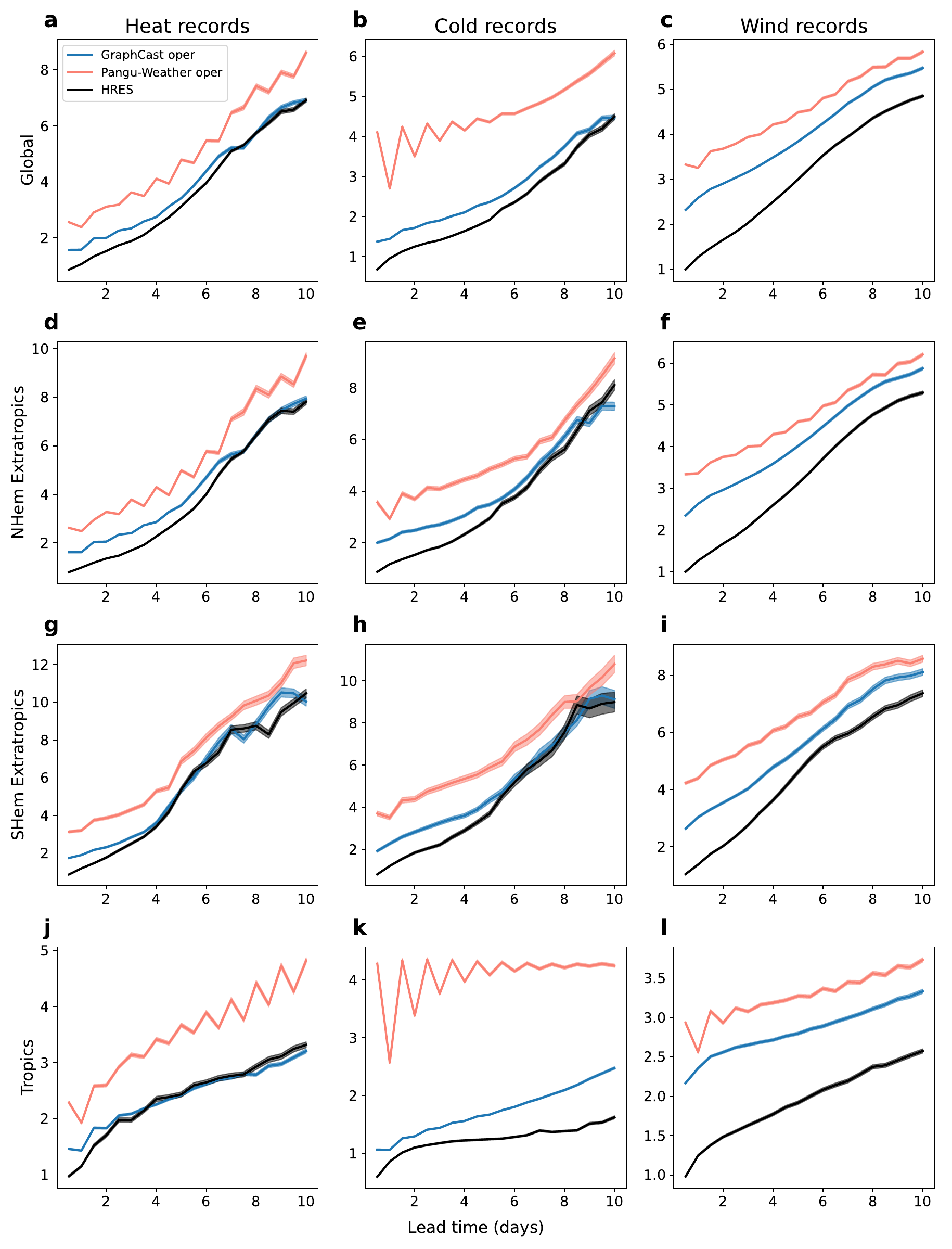}
    \caption{\textbf{Regional RMSE of 2m temperature and 10m wind speed over land for operational forecasts in 2020}. 
    \textbf{a}--\textbf{c}, RMSE for the whole globe.
    \textbf{d}--\textbf{f}, RMSE for Northern Hemisphere Extratropics (NHem Extratropics, i.e., grid cells with latitude in $[20, 90]$).
    \textbf{g}--\textbf{i}, RMSE for Southern Hemisphere Extratropics (SHem Extratropics, i.e., grid cells with latitude in $[-90, -20]$).
    \textbf{j}--\textbf{l}, RMSE for Tropics (grid cells with latitude in $[-20, 20]$).
    The transparent shaded areas indicate $95\%$ confidence bands. 
    }
    \label{figure::RMSE_regional_2020_operational}
\end{figure}

\begin{figure}[H] 
\renewcommand\figurename{Supplementary Fig.} 
    \centering
    \includegraphics[width=\textwidth]{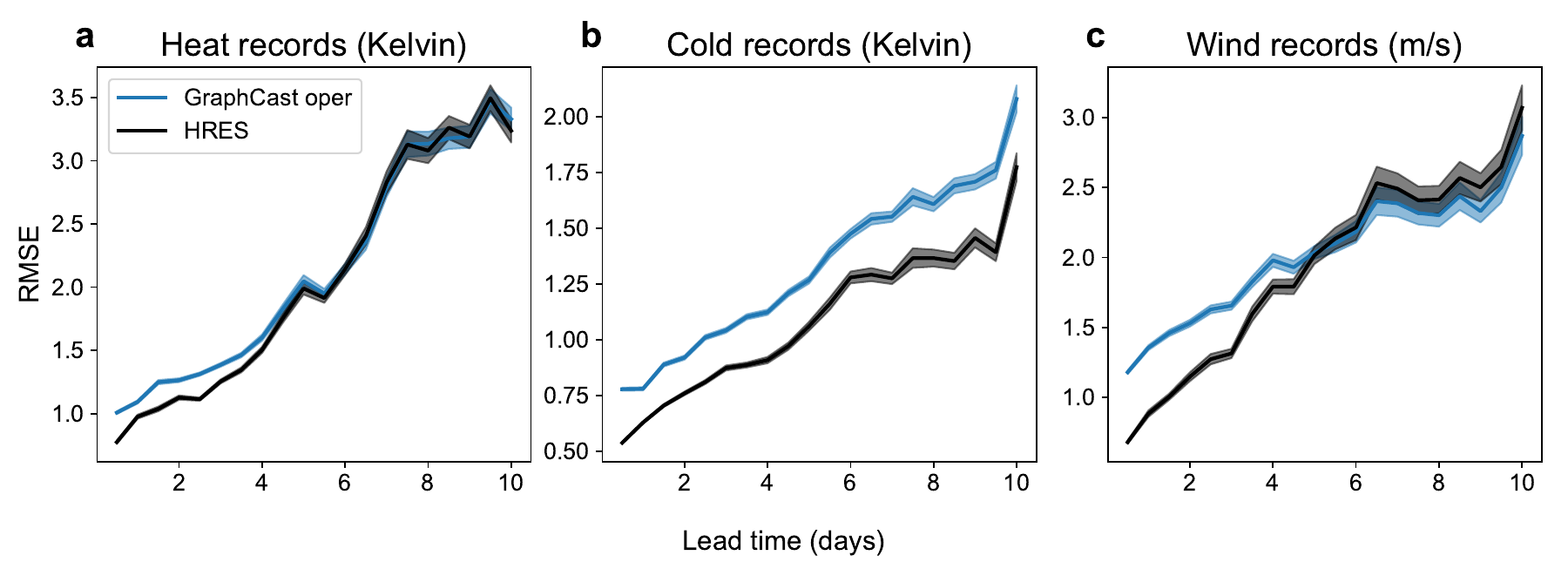}
    \caption{\textbf{RMSE when conditioning on the forecasts rather than the observational ground truth.} RMSE of heat (\textbf{a}), cold (\textbf{b}), and wind records (\textbf{c}) over land (excluding the Antarctic region) in 2020, when the record-breaking events are selected using the HRES and GraphCast operational forecasts (when both of them issue a record-breaking forecast).
    The transparent shaded areas indicate $95\%$ confidence bands. 
    }
    \label{figure::RMSE_forecast_bias_2020_oper_cond_HRES_GC}
\end{figure}

\section{Supplementary figures for the evaluation of non-operational forecasts in 2018}

\begin{figure}[H]
\renewcommand\figurename{Supplementary Fig.} 
    \centering
    \includegraphics[width=\textwidth]{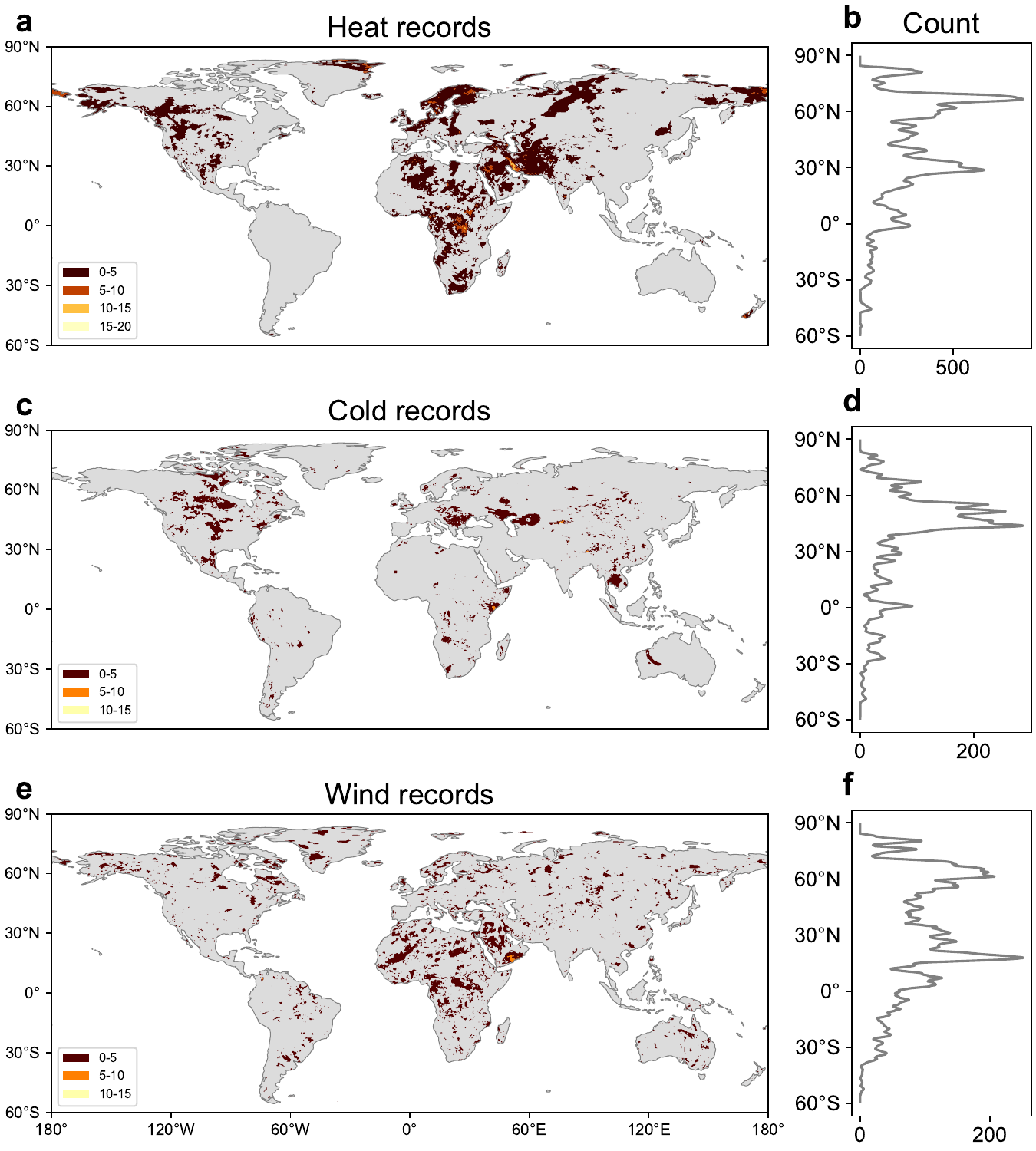}
    \caption{\textbf{Number of records over land (excluding the Antarctic region) in 2018 in ERA5}. \textbf{a}, \textbf{c}, \textbf{e}, Number of heat, cold, and wind records. \textbf{b}, \textbf{d}, and \textbf{f}, Number of heat, cold, and wind records per latitude.}
    \label{figure::map_records_ERA5_2018}
\end{figure}

\begin{figure}[H] 
\renewcommand\figurename{Supplementary Fig.} 
    \centering
    \includegraphics[width=\textwidth]{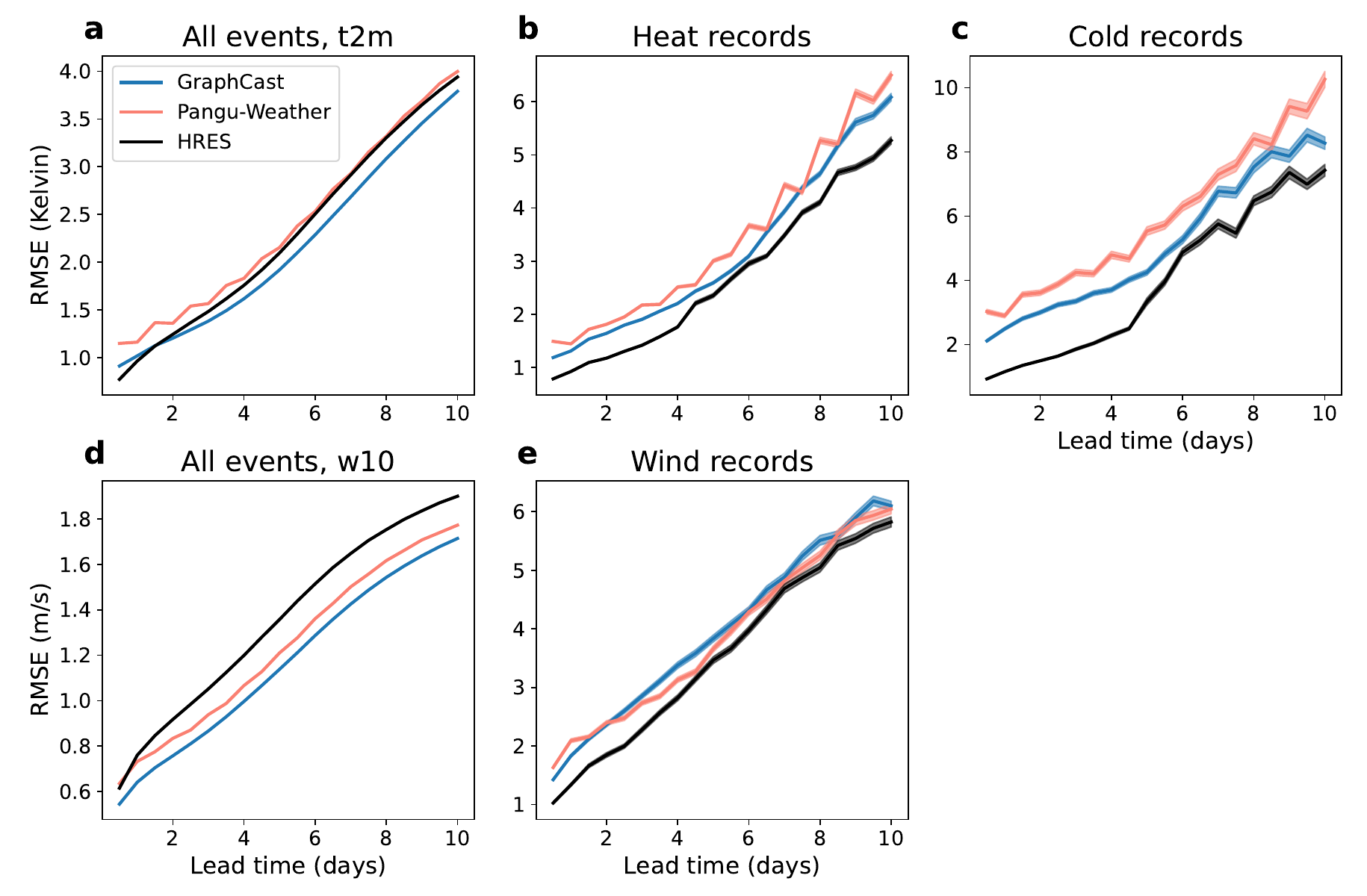}
    \caption{\textbf{Model performance on all events and record-breaking events in 2018}. \textbf{a}--\textbf{e}, RMSE of 2m temperature and 10m wind speed over land (excluding the Antarctic region) of HRES, Pangu-Weather, and GraphCast for all events (\textbf{a}, \textbf{d}) and only record-breaking events (\textbf{b}, \textbf{c}, \textbf{e}) in 2018. The transparent shaded areas indicate $95\%$ confidence bands.
    }
    \label{figure::RMSE_t2m_w10_2018}
\end{figure}

\begin{figure}[H] 
\renewcommand\figurename{Supplementary Fig.} 
    \centering
    \includegraphics[width=\textwidth]{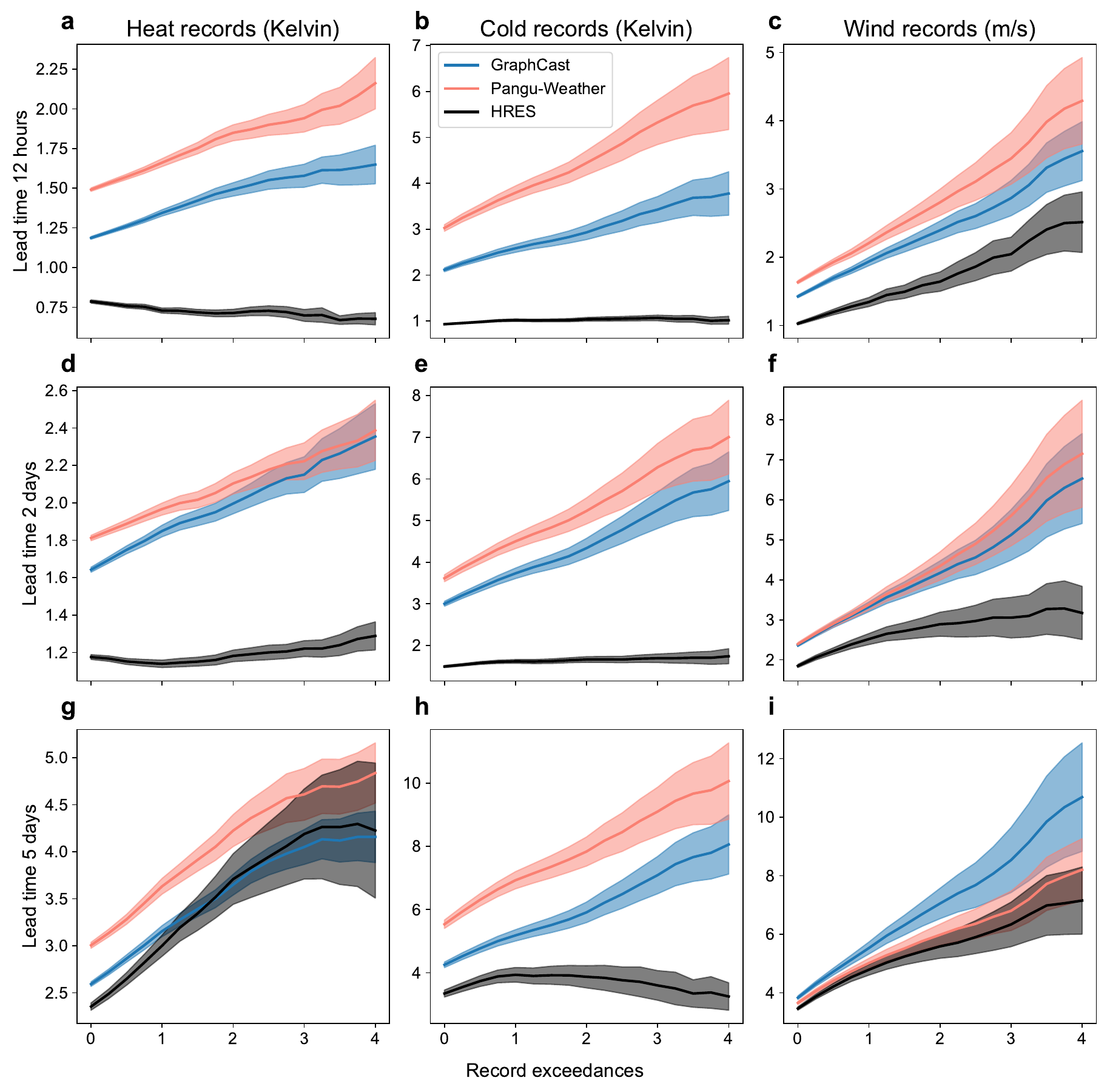}
    \caption{\textbf{RMSE against record exceedances in 2018}. RMSE of 2m temperature and 10m wind speed over land (excluding the Antarctic region) of HRES, Pangu-Weather, and GraphCast in 2018, for events that exceed the record by at least a certain margin, for lead time 12 hours (\textbf{a}--\textbf{c}), 2 days (\textbf{d}--\textbf{f}), and 5 days (\textbf{g}--\textbf{i}). 
    The transparent shaded areas indicate $95\%$ confidence bands.
    }
    \label{figure::RMSE_record_exceedance_2018}
\end{figure}

\begin{figure}[H]
\renewcommand\figurename{Supplementary Fig.} 
    \centering
    \includegraphics[width=\textwidth]{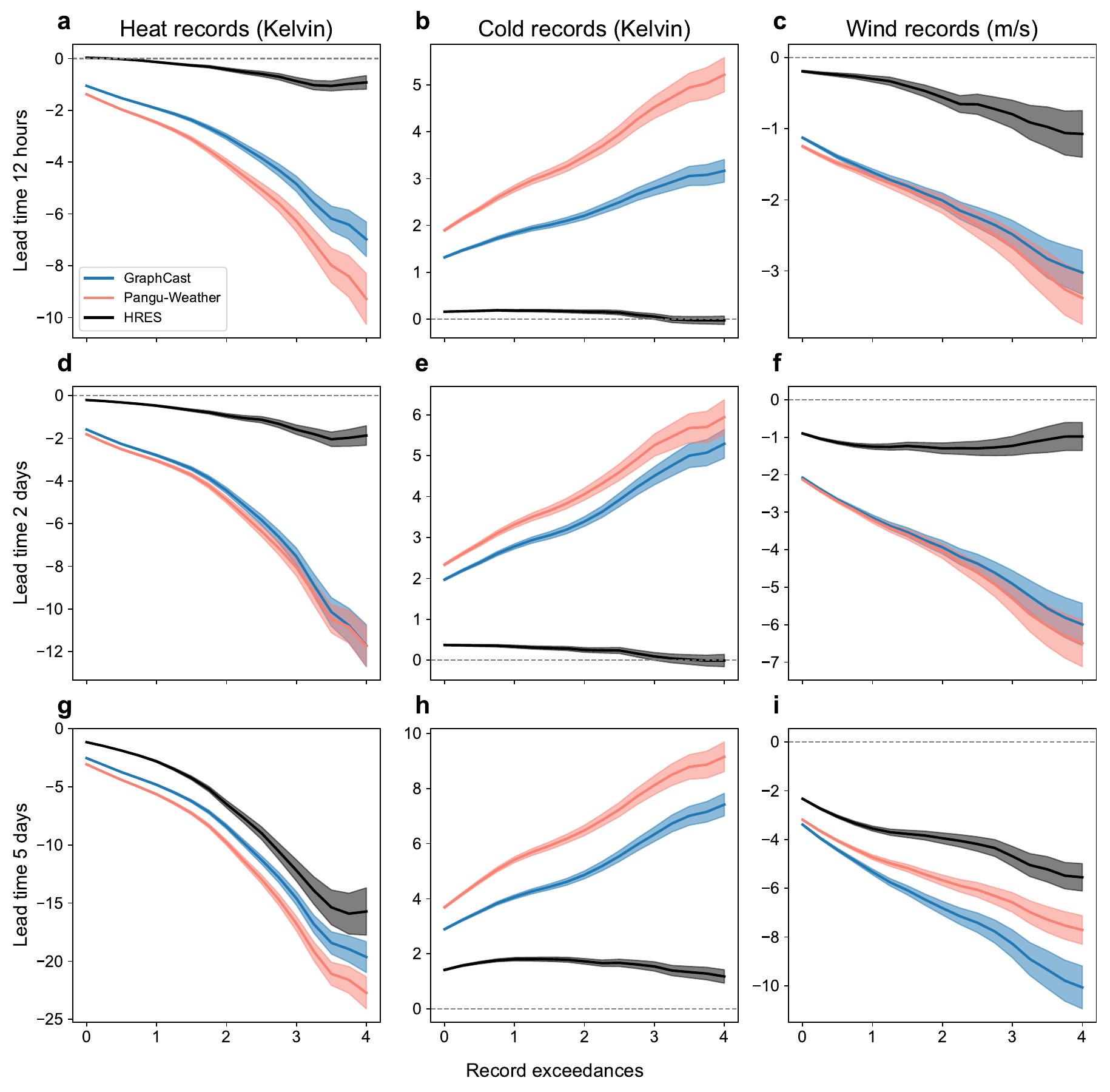}
    \caption{\textbf{Forecast bias against record exceedances in 2018}. Forecast bias of 2m temperature and 10m wind speed over land (excluding the Antarctic region) of HRES, Pangu-Weather, and GraphCast in 2018, for events that exceed the record by at least a certain margin, for lead time 12 hours (\textbf{a}--\textbf{c}), 2 days (\textbf{d}--\textbf{f}), and 5 days (\textbf{g}--\textbf{i}). 
    }
    \label{figure::FB_record_exceedance_2018}
\end{figure}

\begin{figure}[H]
\renewcommand\figurename{Supplementary Fig.} 
    \centering
    \includegraphics[width=0.95\textwidth]{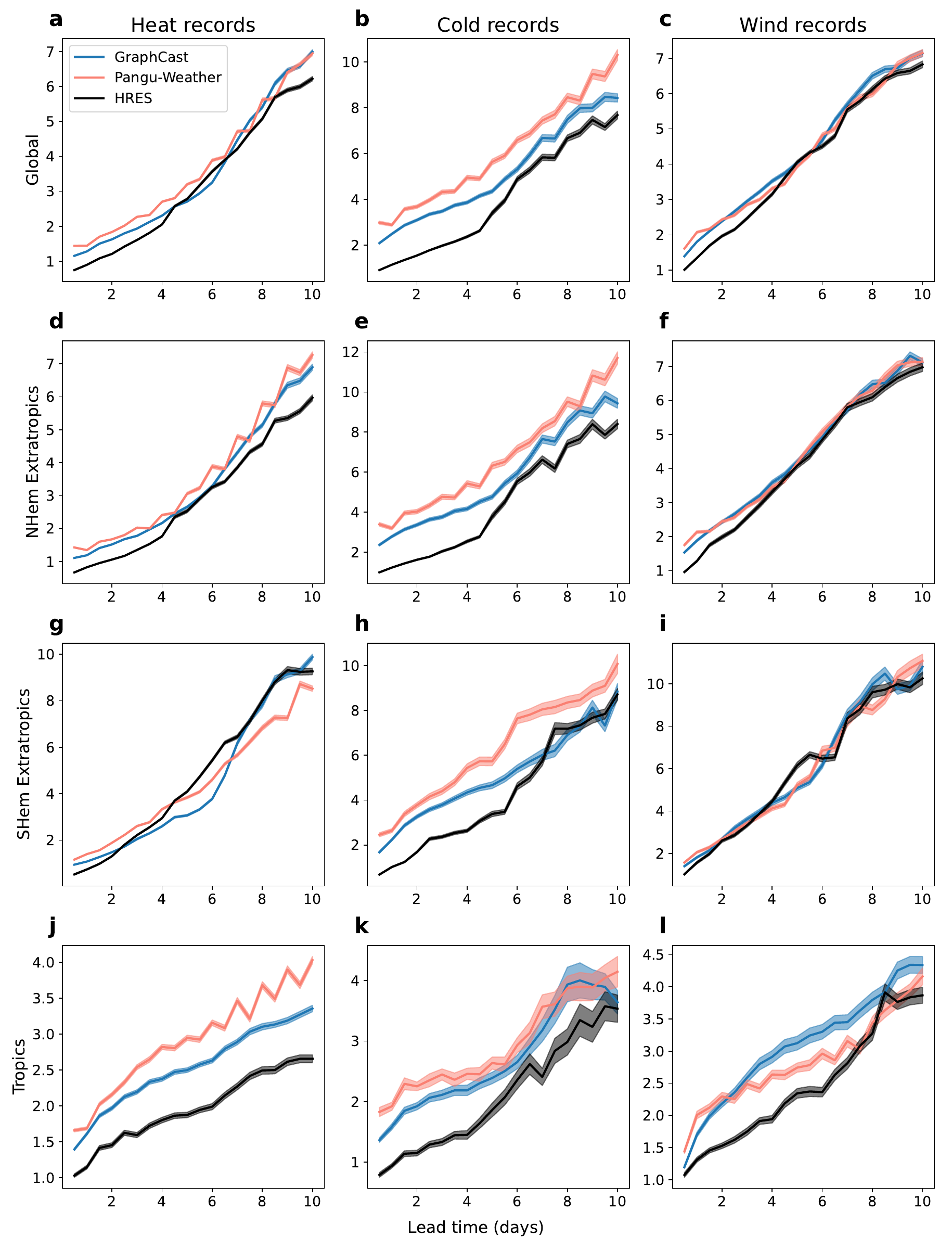}
    \caption{\textbf{Regional RMSE of 2m temperature and 10m wind speed over land in 2018}. 
    \textbf{a}--\textbf{c}, RMSE for the whole globe.
    \textbf{d}--\textbf{f}, RMSE for Northern Hemisphere Extratropics (NHem Extratropics, i.e., grid cells with latitude in $[20, 90]$).
    \textbf{g}--\textbf{i}, RMSE for Southern Hemisphere Extratropics (SHem Extratropics, i.e., grid cells with latitude in $[-90, -20]$).
    \textbf{j}--\textbf{l}, RMSE for Tropics (grid cells with latitude in $[-20, 20]$).
    The transparent shaded areas indicate $95\%$ confidence bands. 
    }
    \label{figure::RMSE_regional_2018}
\end{figure}

\end{document}